\documentclass[12pt]{article}
\usepackage{amssymb}
\usepackage{amsmath}
\usepackage{geometry}
\usepackage[onehalfspacing]{setspace}
\usepackage[pagewise,left,switch]{lineno}

\newtheorem{theorem}{Theorem}

\newtheorem{claim}[theorem]{Claim}

\newtheorem{definition}{Definition}

\newtheorem{proposition}{Proposition}
\newtheorem{remark}{Remark}

\input{tcilatex}
\begin{document}

\title{Local dominance\thanks{%
Previous versions of this paper appeared on Arxiv:
https://arxiv.org/abs/2012.14432v1. \ \ \ \ \ \ \ \ \ \ \ We thank Pierpaolo
Battigalli, In\'{a}cio B\'{o}, Shurojit Chatterji, Yi-Chun Chen, Atsushi
Kajii, Takashi Kunimoto, Jiangtao Li, Alexander Nesterov, Antonio Penta,
Sergey Stepanov, and Satoru Takayashi. We also want to thank all the
attendants of our presentations at SAET 2022, TARK 2021, the 2021 Asian
Meeting of the Econometric Society, and the International Conference on
Social Choice \& Voting Theory, as well as of seminars at the School of
Economics Research Workshop (Singapore Management University), the Institute
of Economics (Academia Sinica), the CCBEF Seminar Series (Southwestern
University of Finance and Economics), and the Advanced Research Forum in
Economics (Economics and Management School, Wuhan University).}}
\author{Emiliano Catonini\medskip \thanks{%
New York University in Shanghai, emiliano.catonini@nyu.edu} and Jingyi Xue%
\thanks{%
Singapore Management University, School of Economics, jyxue@smu.edu.sg.}}
\maketitle

\begin{abstract}
We define notions of dominance between two actions in a dynamic game. Local
dominance considers players who have a blurred view of the future and
compare the two actions by first focusing on the outcomes that may realize
at the current stage. When considering the possibility that the game may
continue, they can only check that the local comparison is not overturned
under the assumption of "continuing in the same way" after the two actions
(in a newly defined sense). Despite the lack of forward planning, local
dominance solves dynamic mechanisms that were found easy to play and
implements social choice functions that cannot be implemented in
obviously-dominant strategies.

\end{abstract}

\section{Introduction}

\begin{quotation}
\textit{The point of view under discussion may be symbolized by the proverb
\textquotedblleft Look before you leap\textquotedblright , and the one to
which it is opposed by the proverb \textquotedblleft You can cross that
bridge when you come to it\textquotedblright . [...] To cross one's bridges
when one comes to them means to attack relatively simple problems of
decisions by artificially confining attention to so small a world that the
principle of \textquotedblleft Look before you leap\textquotedblright\ can
be applied there. I am unable to formulate criteria for selecting these
small worlds and indeed believe that their selection may be a matter of
judgement and experience about which it is impossible to enunciate complete
and sharply defined general principles. (Savage, 1954)}
\end{quotation}

\bigskip

Traditionally, game theory has treated any given game as a small world in
which players can scrutinize all possible contingencies and anticipate all
their future decisions. But empirical and experimental evidence have shown
that this accurate analysis of the game may be too hard for real players.\
Li (2017) addressed players' difficulties with contingent reasoning by
introducing \emph{obvious dominance}:\ a strategy obviously dominates
another strategy if, at each decision node from which the two strategies
depart, the worst outcome that is still possible under the first strategy is
not worse than the best outcome that is still possible under the second
strategy. With this, obvious dominance explores the idea that dynamic
mechanisms may be easier to play because they allow to compare strategies
under smaller outcome sets. Pycia and Troyan (2022) further address players'
difficulties with planning by introducing \emph{strong obvious dominance}:
an action is strongly obviously dominant at a decision node if the worst
outcome that may follow the first action is not worse than the best outcome
that may follow any other action. Thus, strong obvious dominance explores
the idea that players compare actions without a forward plan.

In this paper, we take the following perspective on players' choices in
dynamic games. Instead of planning at the outset, players tackle one choice
problem at a time --- they \textquotedblleft cross a bridge when they come
to it\textquotedblright\ --- and compare the available actions by first
focusing on their \emph{possible immediate consequences}. At the same time,
players realize that, under some actions, the game may continue --- they
\textquotedblleft look before they leap\textquotedblright\ ---, but their
view of the continuation game is blurred, so they only look for confirmation
of the local comparison through simple considerations about the future. We
introduce a notion of \emph{local dominance} between two actions that
captures this perspective on the game.

To illustrate, consider the \textquotedblleft Japanese\textquotedblright\
version\ of a single-unit, ascending-price auction with a discrete clock. At
each stage, players simultaneously choose between \textquotedblleft
leaving\textquotedblright\ and \textquotedblleft bidding\textquotedblright .
The object is assigned at the current price either when one player bids, and
then she is the winner, or when no player bids, in which case the winner is
determined at random. Take the viewpoint of a player who values the object
above the current price. Bidding may immediately yield the object for sure,
if no one else bids, or let the auction continue, otherwise. Before
considering the latter scenario, which requires foresight, it may be natural
to compare the possible, immediate outcomes of the two actions in the former
scenario. Since leaving only yields the object with some probability,
bidding beats leaving. This comparison is simple because it only involves
the few outcomes associated with the current price. By contrast, in a
sealed-bid, second-price auction, each bid can result in winning the object
at many different prices. In general, comparing actions in a static
mechanism may be difficult because each action can immediately induce a
plethora of outcomes. Dynamic mechanisms, instead, draw players' attention
to the few outcomes that can realize at each stage. Local dominance captures
this simplicity factor of dynamic mechanisms. By contrast, obvious and
strong obvious dominance drag into the same picture present and future
outcomes of a strategy/action. As a consequence, in the ascending auction,
bidding until the price reaches one's valuation is not obviously dominant,
because it may result in eventually losing the object, while leaving earlier
may result in winning the object at the lottery. Local dominance, instead,
focuses first on the possible \emph{immediate} outcomes of \emph{both}
actions. This is a coarse form of contingent reasoning, arising from the
separate treatment of present and future of the game in the player's mind.

Players might base their choices\emph{\ only} on their possible immediate
consequences and just ignore the continuation game as too complicated to
analyze.\footnote{%
Our positive results for dynamic mechanisms would obviously hold through if
local dominance was defined with just the comparison of the outcomes that
can realise when both actions terminate the game.} However, they may also be
concerned that the local comparison would be overturned if the game
continues, in ways they cannot fully scrutinize. Because of this concern,
such players may even settle an action that ranks clearly below another
action in terms of possible immediate outcomes.\footnote{%
For instance, in the dynamic translation of the Top Trading Cycles
allocation rule where players demand one of the still-available objects at a
time, a player may be tempted to ask for a less-preferred object if she
fears that the opportunity to obtain it will fade and the opportunity to
obtain a more-preferred one will never arise. We will address this concern
with specific game rules and local dominance in Section 5.2.} Local
dominance ensures that players can find confirmation of the local
comparison\ without forward planning, and yet without careful scrutiny of 
\emph{all} possible future outcomes.

Consider first the possibility that the game continues after the candidate
dominating action but not after the alternative. In the auction, this occurs
when not just our player but also some opponent bids. In this scenario,
while leaving implies losing the object, the final outcome after bidding
will depend on the future moves. However, our player does not reason about
how she will actually play. Instead, she only entertains the idea of leaving
at the next stage, without coming up with any creative alternative. By doing
so, she excludes the possibility of a loss after bidding, confirming its
superiority to leaving. But what makes \textquotedblleft leaving at the next
stage\textquotedblright\ a salient continuation strategy? In our view, the
fact that it \textquotedblleft mimics\textquotedblright\ the alternative
under consideration (in a sense we will make precise). Local dominance will
endogenize in this way what continuation strategy is simple for the player,
requiring that it\ is available and that it confirms the local comparison.

In other situations, the game may continue after both actions under
comparison. To give an example, add action \textquotedblleft
wait\textquotedblright\ to the ascending-price auction. Waiting differs from
leaving only in that, if the auction continues, a player who waited can
still move. The opponents will not observe whether our player waited or bid.
But then, which of the two actions she chose makes no difference whatsover
for the future: she can play in the same way after the two actions, and then
she will also get the same outcome. Local dominance will thus require that
the scenario in which the game continues after both actions is irrelevant
for the choice and can be rightfully ignored. Realising that the present
choice cannot have any impact on the continuation game does not truly
require to scrutinize all possible ways in which the game may continue, but
just a rough understanding of the game pattern. Recognizing the invariance
of future outcomes could also be easier than finding best and worst outcomes
across two heterogenous outcome sets, because it does not require to
understand the details of a possibly complicated outcome rule.\footnote{%
In Section 6.3, we consider an example of a maze in which, at the first
bifurcation, after going right the player may or may not find the exit,
whereas after going left she will certainly not find it, but discovering
this requires detailed scrutiny of two very complicated subtrees.}

In static game, local and (strong)\ obvious dominance coincide. But in a
dynamic game, even a strongly obviously dominant action need not be locally
dominant. Yet, local dominance provides a possible explanation of why some
dynamic mechanisms that are not obviously strategy-proof were found easy to
play, and it yields positive implementation results for relevant social
choice functions that cannot be implemented in obviously dominant
strategies. The \textquotedblleft Japanese\textquotedblright\
ascending-price auction (with the random tie-breaking rule) was found easy
to play in the experimental work of Kagel et al. (1987). Moreover, we
construct a dynamic mechanism that implements the Top Trading Cycles
allocation rule in locally dominant actions, although (as shown by Li, 2017)
no obviously strategy-proof mechanism can implement it.\footnote{%
Our mechanism is a special case of the class of \textquotedblleft menu
mechanisms\textquotedblright\ defined by Mackenzie and Zhou (2022). In
particular, it is very similar to Bo and Hakimov's (2022) \textquotedblleft
pick-an-object mechanism\textquotedblright\ for the implementation of the
TTC\ rule, the difference being in players' information flow. Bo and Hakimov
(2022) also provide experimental evidence for the simplicity of their
mechanism. See Section 6.3 for details.}

\bigskip

Local dominance builds on a more general approach that we develop, whereby
players compare actions rather than strategies, under a partition of the
contingencies and without a plan for the future. We start with an
exogenously given partition of the possible states of nature and opponents'
strategies, and we call each partition element a \textquotedblleft
scenario\textquotedblright . We say that action $a$ \emph{dominates} action $%
b$ given the partition if, for every continuation strategy after $b$, there
is a continuation strategy after $a$ that guarantees a better outcome in
every scenario. A distinctive feature of our approach is that our player
does not associate the candidate dominating action with \emph{one}
continuation strategy --- continuation strategies are only used as mental
checks and hence can change with the alternative under consideration.
Coherently with this idea, we further allow the flexibility of tailoring the
continuation strategy on the scenario under analysis. We say that action $a$ 
\emph{scenario-by-scenario dominates} (s-dominates) action $b$ if, in each
scenario, for every continuation strategy after $b$, there is a continuation
strategy after $a$ that can only do better. Thus, to establish s-dominance,
each scenario is analyzed in isolation.\ Because of this, in each scenario,
s-dominance does not truly require to scrutinize the possible continuation
strategies after $b$; one can just look for the best possible outcome after $%
b$ and compare it with the worst possible outcome after $a$ under some
good-enough continuation strategy.

Dominance spans between the case of perfect contingent reasoning (with the
finest partition of the space of uncertainty) and the case of no contingent
reasoning (with the coarsest partition). In the first case, we talk of \emph{%
weak dominance} between actions, in the second case of \emph{obvious
dominance}. Because of the lack of global planning, by choosing
weakly/obviously \emph{un}dominated actions a player may follow a
weakly/obviously dominated strategy --- a form of dynamic inconsistency.
Yet, we show that a strategy is weakly/obviously dominant if and only if it
prescribes a weakly/obviously dominant action at every information sets that
can be reached when playing that strategy. Thus, in a (obviously)
strategy-proof dynamic game, a player does not have to recognize the
existence of the dominant strategy in advance; by just spotting one dominant
action at a time, she will --- perhaps unknowingly --- carry it out. In
other words, the flexibility in the use of continuation strategies can be
exploited without making mistakes.

Since a dominant action is s-dominant (given the same partitions), a player
can also discover it without reasoning across scenarios, hence using the
most convenient continuation strategy in each scenario. Perhaps
surprisingly, we show that the following converse implication also holds:
when there is an s-dominant action at every information set, such actions
are also dominant. Thus, the simpler approach of s-dominance is equally
effective to determine whether there is a dominant action at each decision
node. With perfect contingent reasoning, s-dominance boils down to comparing
the best possible outcomes after the two actions in each contingency. This
is a form of wishful thinking (we call it \emph{wishful dominance}) that
eliminates the need to even conceive continuation strategies. Seen from this
angle, if players are capable of perfect contingent reasoning, dynamic
strategy-proof mechanisms do not require any sort of planning, and this
could be an explanation for their simplicity.

While a (s-)dominance relation between actions may be easy to spot, a priori
there is no guarantee of this. First, even the coarsest scenarios that allow
to establish dominance may be hard to identify. Second, the continuation
strategies that allow to establish dominance may be hard to identify as
well; the optimal continuation strategy may even be the only one that does
the job. Therefore, with local dominance, we endogenize the scenarios from
the local viewpoint, and we impose a simple use of continuation strategies.

The partition associated with local dominance, which we call
\textquotedblleft local partition\textquotedblright , is driven by the
natural separation in a player's mind between the scenario in which an
action terminates the game, and hence yields an immediate consequence, and
the scenario in which it does not, and hence one must look ahead. With this,
our player identifies the four scenarios in which each of the two actions
either terminates her game or not.\footnote{%
Both in the ascending-price auction (without waiting action)\ and in our
TTC\ mechanism, local dominance only requires to distinguish the two
scenarios in which the \emph{dominant} action terminates the game or not.
This is particularly convenient in the TTC mechanism, because it allows to
use the same unique partition of the contingencies across all comparisons
with many alternatives.} These scenarios are \textquotedblleft
local\textquotedblright , in the sense that they only depend on the moves of
the opponents before the next decision nodes of our player, avoiding any
consideration on their future moves. In terms of continuation strategies, in
the scenarios where the candidate dominating action $a$ terminates the game,
a player does not actually need to entertain any continuation strategy. In
the scenarios where action $a$ does not terminate the game, compared to
s-dominance, local dominance restricts the use of continuation strategies by
only allowing to compare action $a$ with action $b$ under the hypothesis of 
\emph{continuing in the same way}. By doing so, our player defers any
consideration about the optimality of her future moves. In the scenario
where $b$ terminates the game, \textquotedblleft continuing in the same
way\textquotedblright\ after $a$ translates into playing $b$ or an action
that terminates the game --- in the auction, \textquotedblleft
leaving\textquotedblright . In the more complicated scenario where the game
continues after both actions, local dominance departs from s-dominance by
requiring that the final outcome is invariant to the present choice.

With the idea of \textquotedblleft continuing in the same
way\textquotedblright , our player compares the two actions \emph{ceteris
paribus} with respect to all other decisions she will make. This can be seen
as an adaptation of the classical one-shot deviation check for a player who
does not have a plan. More importantly, the idea of continuing in the same
way allows to apply the \emph{sure thing principle} (STP) to the scenario in
which the game continues after both actions. We find this application of
STP\ simple because such a scenario is already isolated in our player's mind
--- we contrast this with a classical violation of STP in Section 4.

\bigskip

The paper is organized as follows. In Section 2 we describe the game. In
Section 3 we introduce and analyze our baseline notion of dominance and the
notion of s-dominance. Section 4 is devoted to local dominance. Section 5 is
devoted to applications of local dominance: we solve the ascending auctions
and design our dynamic TTC mechanism. Section 6 concludes with a final
comparison with Li (2017) and Pycia and Troyan (2022), along with a
discussion of other related literature and an avenue for future research.
The Appendix collects the formal proofs of some of the results and an
example of a relaxation of local strategy-proofness (\emph{on-path
strategy-proofness}) that could be natural and more appropriate in some
contexts.

\section{Framework}

We consider a finite multistage game in which players possess
payoff-relevant private information, which we represent as the asymmetric
observation of an initial move of nature. Nature has a finite set $\Theta $
of possible moves, and each player $i\in I$ has a finite set $A_{i}$ of
actions that are available at some point of the game; let $A=\times _{i\in
I}A_{i}$. Without loss of generality, we assume that, after the move of
nature, the game always lasts exactly $T$ stages, and that all players
simultaneously choose an action at every stage. (When a player is not truly
active, only a dummy action will be available.) Thus, the set $X^{1}$ of
stage-$1$ histories is $\Theta $, and for each $t=2,...,T$, the set $X^{t}$
of stage-$t$ histories is a subset of $\Theta \times A^{t-1}$; the set of
terminal histories $Z$ is a subset of $\Theta \times A^{T}$. Let $X=\cup
_{t=1}^{T}X^{t}$ and $\overline{X}=X\cup Z$; $\overline{X}$ is endowed with
the \textquotedblleft prefix of\textquotedblright\ partial order, denoted by
\textquotedblleft $\prec $\textquotedblright . An information set of player $%
i$ collects histories of the \emph{same} stage that player $i$ cannot
distinguish (players know the stage); the set of $i$'s information sets $%
H_{i}$ partitions $X$ and satisfies perfect recall, and thus it inherits the
precedence order $\prec $ from $\overline{X}$. For each $h\in H_{i}$, let $%
A_{i}^{h}\subseteq A_{i}$ denote the set of actions available to player $i$
at $h$. Let $H_{i}^{\ast }$ collect the information sets $h\in H_{i}$ where
player $i$ is active, that is, $\left\vert A_{i}^{h}\right\vert >1$.

For each player $i$, there is a set of possible outcomes $Y_{i}$. Let $%
g_{i}:Z\rightarrow Y_{i}$ denote the function that associates each terminal
history with $i$'s final outcome, and let $u_{i}:Z\rightarrow 
\mathbb{R}
$ denote player $i$'s payoff function.

A reduced strategy of player $i$ (henceforth, just \textquotedblleft
strategy\textquotedblright ) is a map $s_{i}$ that assigns an action $%
a_{i}\in A_{i}^{h}$ to each information set $h\in H_{i}$ that can be reached
given the actions assigned to the previous information sets. Note that a
strategy prescribes a dummy action also at every information set where
player $i$ is not active. Let $S_{i}$ denote the set of strategies of player 
$i$, and let $S_{-i}=\Theta \times (\times _{j\in I\backslash \left\{
i\right\} }S_{j})$. For each $(s_{i,}s_{-i})\in S_{i}\times S_{-i}$, let $%
\zeta (s_{i,}s_{-i})$ denote the induced terminal history. For each $h\in
H_{i}$, let $S_{i}(h)$ and $S_{-i}(h)$ denote, respectively, the elements of 
$S_{i}$ and $S_{-i}$ that allow to reach $h$. For each $a_{i}\in A_{i}^{h}$,
let $S_{i}(h,a_{i})$ denote the set of strategies $s_{i}\in S_{i}(h)$ such
that $s_{i}(h)=a_{i}$. Although, formally, an element of $S_{i}(h,a_{i})$ is
a strategy for the entire game, we will use $S_{i}(h,a_{i})$ to describe the 
\emph{continuation strategies} of $i$ after choosing $a_{i}$ at $h$. Let $%
H_{i}^{\ast }(s_{i})$ denote the set of active information sets of $i$ that
are consistent with strategy $s_{i}$, i.e., that can be reached by playing $%
s_{i}$.

\section{Dominance between actions}

\subsection{Baseline notion of dominance}

We take the viewpoint of a player who compares two available actions $%
\overline{a}_{i},a_{i}$ at an information set $h\in H_{i}^{\ast }$. As we
are mainly interested in the existence of a dominant action, we only
consider comparisons between pure actions. The relevant state for player $i$%
's choice is $s_{-i}$, which includes the move of nature $\theta $ and the
strategies of the opponents $(s_{j})_{j\not=i}$.\footnote{%
For coherence with the view that player $i$ does not formulate a global
plan, the strategies of the opponents are best interpreted not as plans
either, but just as sets of conditional statements about behavior:\ how a
player would behave upon reaching an information set.} Given the current
information set $h$, the set $S_{-i}(h)$ is thus the set of relevant states
that are still possible at $h$. A \emph{scenario }is a subset of $S_{-i}(h)$
under which player $i$ compares the two actions. We assume that player $i$
considers a collection of scenarios that partitions $S_{-i}(h)$. For
example, a player could ask herself \textquotedblleft will my opponent go
left or right at the current stage?\textquotedblright , and compare actions
under each of the two scenarios of the corresponding bipartition. Our
baseline notion of dominance postulates how a player compares the two
actions under a given partition.

\begin{definition}
Fix an information set $h\in H_{i}^{\ast }$, an action pair $\left( 
\overline{a}_{i},a_{i}\right) \in A_{i}^{h}\times A_{i}^{h}$, and a
partition $\mathcal{S}$ of $S_{-i}(h)$. Action $\overline{a}_{i}$ \textbf{%
dominates} action $a_{i}$ given $\mathcal{S}$ if for every $s_{i}\in
S_{i}(h,a_{i})$, there exists $\overline{s}_{i}\in S_{i}(h,\overline{a}_{i})$
such that%
\begin{equation}
\forall \overline{S}_{-i}\in \mathcal{S},\text{ \ \ }\min_{s_{-i}\in 
\overline{S}_{-i}}u_{i}\left( \zeta (\overline{s}_{i},s_{-i})\right) \geq
\max_{s_{-i}\in \overline{S}_{-i}}u_{i}\left( \zeta (s_{i},s_{-i})\right) .
\label{LD1}
\end{equation}%
Action $\overline{a}_{i}$ is dominant (at $h$) if it dominates every other $%
a_{i}\in A_{i}^{h}$ given the partition associated with $\left( \overline{a}%
_{i},a_{i}\right) $.
\end{definition}

Action $\overline{a}_{i}$ dominates action $a_{i}$ if for every possible
continuation strategy after $a_{i}$, there exists a continuation strategy
after $\overline{a}_{i}$ that does better in every scenario. Local dominance
does not require player $i$ to find \emph{one} continuation strategy after $%
\overline{a}_{i}$ that beats \emph{all} continuation strategies after $a_{i}$
--- such a continuation strategy may not even exist. Rather, the
continuation strategy after $\overline{a}_{i}$ conceived by player $i$ can
change with the continuation strategy after $a_{i}$ that player $i$ is
considering. As we will see in Section \ref{Section: very local dominance},
this is particularly convenient when for every continuation strategy after $%
a_{i}$, there is a \textquotedblleft mimicking\textquotedblright\
continuation strategy after $\overline{a}_{i}$ which does the job. Our
baseline notion of dominance does not allow, instead, to change the
continuation strategy after $\overline{a}_{i}$ depending on the scenario
under consideration --- we will formalize a notion of dominance with this
additional flexibility in the next subsection.

To establish that a continuation strategy after $\overline{a}_{i}$
\textquotedblleft does better\textquotedblright\ than a continuation
strategy after $a_{i}$ within a scenario, we employ the comparison of the
worst and the best outcome that can realize under, respectively, the former
and the latter continuation strategies. This conservative approach, borrowed
from obvious dominance, makes our notions of dominance robust to possible
ways in which players compare actions within a scenario.

It is easy to see that the finer is the partition into scenarios, the weaker
is the notion of dominance associated with the partition.

\begin{remark}
\label{Remark: partitions}Fix two partitions $\mathcal{S},\widetilde{%
\mathcal{S}}$ of $S_{-i}(h)$ where $\widetilde{\mathcal{S}}$ refines $%
\mathcal{S}$. If $\overline{a}_{i}$ dominates $a_{i}$ given $\mathcal{S}$,
then $\overline{a}_{i}$ dominates $a_{i}$ given $\widetilde{\mathcal{S}}$.
\end{remark}

Consider now the finest and the coarsest partitions of $S_{-i}(h)$: the
singleton partition $\mathcal{S}=\left\{ \left\{ s_{-i}\right\} |s_{-i}\in
S_{-i}(h)\right\} $, and the trivial partition $\mathcal{S}=\left\{
S_{-i}(h)\right\} $. Under these two partitions, in light of Remark \ref%
{Remark: partitions}, we obtain the weakest and the strongest notions of
dominance we can establish by just varying the level of detail in contingent
reasoning. We formalize these two extreme notions of dominance between
actions and call them \textquotedblleft weak dominance\textquotedblright\
and \textquotedblleft obvious dominance\textquotedblright\ because in static
games they coincide with the corresponding notions of dominance between
strategies.\footnote{%
As common in mechanism design but not in game theory, we define weak
dominance without requiring strict inequality under some contingency --- in
game theory, this notion is called \emph{very weak dominance }(Marx and
Swinkels, 1997).}

\begin{definition}
\label{Def: weak/obvious}Fix an information set $h\in H_{i}^{\ast }$.

Action $\overline{a}_{i}\in A_{i}^{h}$ \textbf{weakly dominates} action $%
a_{i}\in A_{i}^{h}$ if for every $s_{i}\in S_{i}(h,a_{i})$, there exists $%
\overline{s}_{i}\in S_{i}(h,\overline{a}_{i})$ such that%
\begin{equation}
\forall s_{-i}\in S_{-i}(h),\text{ \ \ }u_{i}\left( \zeta (\overline{s}%
_{i},s_{-i})\right) \geq u_{i}\left( \zeta (s_{i},s_{-i})\right) .
\label{LWD1}
\end{equation}%
Action $\overline{a}_{i}\in A_{i}^{h}$ \textbf{obviously dominates} action $%
a_{i}\in A_{i}^{h}$ if for every $s_{i}\in S_{i}(h,a_{i})$, there exists $%
\overline{s}_{i}\in S_{i}(h,\overline{a}_{i})$ such that%
\begin{equation}
\min_{s_{-i}\in S_{-i}(h)}u_{i}\left( \zeta (\overline{s}_{i},s_{-i})\right)
\geq \max_{s_{-i}\in S_{-i}(h)}u_{i}\left( \zeta (s_{i},s_{-i})\right) .
\label{LOD1}
\end{equation}%
Action $\overline{a}_{i}$ is weakly/obviously dominant if it
weakly/obviously dominates every other $a_{i}\in A_{i}^{h}$.
\end{definition}

We now compare weak/obvious dominance between actions with\linebreak
weak/obvious dominance between strategies (in dynamic games). We first
report the definitions of the latter. Given two strategies $\overline{s}%
_{i},s_{i}\in S_{i}$, let $\mathcal{D}(\overline{s}_{i},s_{i})$ be the set
of their \emph{points of departure}, that is, the information sets $h\in
H_{i}^{\ast }(\overline{s}_{i})\cap H_{i}^{\ast }(s_{i})$ such that $%
\overline{s}_{i}(h)\not=s_{i}(h)$.\footnote{%
Given our focus on reduced strategies, and thanks to perfect recall, every
point of departure is an \emph{earliest} point of departure in the sense of
Li (2017).}

\begin{definition}
\label{Def: dominance btw strategies}Strategy $\overline{s}_{i}$ weakly
dominates strategy $s_{i}$ if%
\begin{equation}
\forall s_{-i}\in S_{-i},\text{ \ \ }u_{i}\left( \zeta (\overline{s}%
_{i},s_{-i})\right) \geq u_{i}\left( \zeta (s_{i},s_{-i})\right) .
\label{WD1}
\end{equation}%
Strategy $\overline{s}_{i}$ obviously dominates strategy $s_{i}\ $if%
\begin{equation}
\forall h\in \mathcal{D}(\overline{s}_{i},s_{i}),\text{ \ \ }\min_{s_{-i}\in
S_{-i}(h)}u_{i}\left( \zeta (\overline{s}_{i},s_{-i})\right) \geq
\max_{s_{-i}\in S_{-i}(h)}u_{i}\left( \zeta (s_{i},s_{-i})\right) .
\label{OD1}
\end{equation}%
A strategy is weakly/obviously dominant if it weakly/obviously dominates
every other strategy.
\end{definition}

Weak/obvious dominance between two actions at an information set is
equivalent to the following relationship between strategies that reach the
information set:\ every strategy that prescribes the dominated action is
weakly/ obviously dominated by some strategy that prescribes the dominating
action.\footnote{%
Shimoji and Watson (1998) introduce the notion of conditional (strict)
dominance between strategies at an information set. Proposition \ref%
{Proposition foundation} essentially relates weak and obvious dominance
between actions to the weak and obvious counterparts of conditional
dominance.}

\begin{proposition}
\label{Proposition foundation}Action $\overline{a}_{i}\in A_{i}^{h}$
weakly/obviously dominates action $a_{i}\in A_{i}^{h}$ if and only if every $%
s_{i}\in S_{i}(h,a_{i})$ is weakly/obviously dominated by some $\overline{s}%
_{i}\in S_{i}(h,\overline{a}_{i})$.
\end{proposition}

The \textquotedblleft only if\textquotedblright\ part of Proposition \ref%
{Proposition foundation} implies that weak dominance between actions has the
following foundation:\ if action $a_{i}$ is weakly dominated by some action $%
\overline{a}_{i}$ (but not the other way round), then choosing $a_{i}$ is
not optimal under any full-support belief over $S_{-i}(h)$. Moreover, the
\textquotedblleft if\textquotedblright\ part guarantees that action $a_{i}$
is (weakly/obviously) dominated by action $\overline{a}_{i}$ if \emph{every}
strategy that prescribes $a_{i}$ is dominated by a strategy that prescribes $%
\overline{a}_{i}$. Note however that a dominated strategy need not prescribe
a dominated action. It can even be the case that all the actions prescribed
by a dominated strategy are undominated. This is a consequence of the lack
of global planning. As an example, consider the following perfect
information game:%
\begin{equation*}
\begin{array}{cccccccccc}
Ann & \longrightarrow & Bob & \longrightarrow & Ann & \longrightarrow & Bob
& \longrightarrow & (3,3) &  \\ 
\downarrow &  & \downarrow &  & \downarrow &  & \downarrow &  &  & 
\longrightarrow \text{= across} \\ 
(2,0) &  & (0,1) &  & (1,0) &  & (0,1) &  &  & \downarrow \text{= down}%
\end{array}%
\end{equation*}%
The strategy of Ann that prescribes $across$ at the initial history and $%
down $ at history $(across,across)$ is weakly/obviously dominated by
strategy $down$. Nonetheless, it is easy to see that both action $across$ at
the initial history and action $down$ at history $(across,across)$ are not
weakly/obviously dominated.

Thus, when players do not follow a global plan, by choosing undominated
actions they might still end up playing a dominated strategy. As a
consequence, fewer patterns of behavior can be ruled out with dominance
between actions in place of dominance between strategies. In light of this,
it seems harder to achieve \textquotedblleft (obvious)\
strategy-proofness\textquotedblright\ without assuming global planning. But
this is not true: when a player has a weakly/obviously dominant strategy,
all the actions it prescribes are weakly/obviously dominant, so weak/obvious
dominance between actions does rule out any other behavior. Thus, we obtain
the following characterization.

\begin{theorem}
\label{CHAR 1}A strategy $\overline{s}_{i}$ is weakly/obviously dominant if
and only if $\overline{s}_{i}(h)$ is\linebreak weakly/obviously dominant at
every $h\in H_{i}^{\ast }(\overline{s}_{i})$.
\end{theorem}

Theorem \ref{CHAR 1} implies that (obvious) strategy-proofness can be broken
down into a collection of local conditions, each requiring the existence of
a\linebreak weakly/obviously dominant action. This local property may be key
to understand why some (obviously)\ strategy-proof dynamic mechanisms are
easier to play than their static, direct counterparts. Dynamic mechanisms
decompose the problem of revealing players' entire preferences into a series
of smaller partial-revelation problems, but this advantage is lost if all
such problems must be jointly solved at the outset --- to what extent can
players really tackle them in isolation? Theorem \ref{CHAR 1} says that
players do not necessarily have to plan globally if the decomposition
preserves the existence of a dominant strategy; instead, they can find one
dominant action at a time. We build our simplicity theory precisely on this
ground: with local dominance we will identify a natural way of tackling each
choice problem in isolation.

\subsection{Scenario-by-scenario dominance}

Now we explore the possibility that, for the comparison of two actions,
players analyze each scenario in isolation. This means that players may use
different continuation strategies while considering different scenarios.

\begin{definition}
\label{Definition S-dominance}Fix an information set $h\in H_{i}^{\ast }$,
an action pair $\left( \overline{a}_{i},a_{i}\right) \in A_{i}^{h}\times
A_{i}^{h}$, and a partition $\mathcal{S}$ of $S_{-i}(h)$. Action $\overline{a%
}_{i}$ scenario--by-scenario dominates (s-dominates) action $a_{i}$ given $%
\mathcal{S}$ if, for every $\overline{S}_{-i}\in \mathcal{S}$,%
\begin{equation}
\forall s_{i}\in S_{i}(h,a_{i}),\text{\ }\exists \overline{s}_{i}\in S_{i}(h,%
\overline{a}_{i}),\text{ }\min_{s_{-i}\in \overline{S}_{-i}}u_{i}\left(
\zeta (\overline{s}_{i},s_{-i})\right) \geq \max_{s_{-i}\in \overline{S}%
_{-i}}u_{i}\left( \zeta (s_{i},s_{-i})\right) .  \label{LCWD1}
\end{equation}%
Action $\overline{a}_{i}$ is s-dominant if it s-dominates every other $%
a_{i}\in A_{i}^{h}$ given the partition associated with $\left( \overline{a}%
_{i},a_{i}\right) $.
\end{definition}

As for the baseline notion of dominance, the finer the partition, the weaker
s-dominance. Given the same partition, s-dominance is weaker than dominance,
because the continuation strategy after the s-dominating action can change
with the scenario. When the partition is trivial s-dominance and dominance
coincide. The following remark summarizes these observations.

\begin{remark}
\begin{enumerate}
\item For any two partitions $\mathcal{S},\widetilde{\mathcal{S}}$ of $%
S_{-i}(h)$ such that $\widetilde{\mathcal{S}}$ refines $\mathcal{S}$, if $%
\overline{a}_{i}$ s-dominates $a_{i}$ given $\mathcal{S}$, then $\overline{a}%
_{i}$ s-dominates $a_{i}$ given $\widetilde{\mathcal{S}}$.

\item If action $\overline{a}_{i}$ dominates action $a_{i}$ given $\mathcal{S%
}$, then $\overline{a}_{i}$ s-dominates $a_{i}$ given $\mathcal{S}$.

\item Action $\overline{a}_{i}$ obviously dominates $a_{i}$ if and only if
it s-dominates $a_{i}$ given the trivial partition $\mathcal{S}=\left\{
S_{-i}(h)\right\} $.
\end{enumerate}
\end{remark}

Since a dominant action is also s-dominant, to discover a dominant action a
player can also analyze every scenario in isolation. Typically, this
simplifies her task, because it allows to tailor the continuation strategy
after the dominant action on the scenario under consideration.\footnote{%
Whether players actually reason in this way is a question for
experimental/empirical research.} Perhaps surprisingly, when there are
s-dominant actions everywhere, the converse implication holds as well: the
s-dominant actions are also dominant. Thus, the simpler approach of
s-dominance is equally effective to determine whether there is a dominant
action at each decision node.

\begin{theorem}
\label{Theorem local dominant = local c-dominant}For each strategy $%
\overline{s}_{i}\in S_{i}$, $\overline{s}_{i}(h)$ is dominant at every $h\in
H_{i}^{\ast }(\overline{s}_{i})$ if and only if $\overline{s}_{i}(h)$ is
s-dominant at every $h\in H_{i}^{\ast }(\overline{s}_{i})$.
\end{theorem}

Of course, Theorem \ref{Theorem local dominant = local c-dominant} assumes
that, for every information set and every alternative action, dominance and
s-dominance are established under the same partition. Instead, surprisingly,
Theorem \ref{Theorem local dominant = local c-dominant} does not require any
discipline across the partitions used at different information sets. One
could expect that, to guarantee that an s-dominant action is dominant, the
existence of a dominant action at every future information set must be
established under sufficiently coarse partitions. This is not the case.

Definition \ref{Definition S-dominance} highlights the key difference
between s-dominance and dominance by simply changing the order of
quantifiers: first the scenario is fixed, then different continuation
strategies are considered. But\ s-dominance can be checked without actually
scrutinizing all possible continuation strategies after the dominated
action. Given a scenario, among the continuation strategies after $\overline{%
a}_{i}$ that verify condition (\ref{LCWD1}) against the continuation
strategies after $a_{i}$, at least one works against \emph{all} the
continuation strategies after $a_{i}$: the one that gives the highest
worst-case payoff within the scenario. But then, condition (\ref{LCWD1})
boils down to checking that such a payoff is larger than the best possible
payoff after $a_{i}$. This yields a convenient operational definition of
s-dominance. To formalize, let $Z(h,a_{i},\overline{S}_{-i})$ denote the
terminal histories that can be reached under $\overline{S}_{-i}\subseteq
S_{-i}(h)$ if $i$ chooses $a_{i}$ at $h$.

\begin{remark}
\label{Remark: simplification local c-dom}Action $\overline{a}_{i}$
s-dominates action $a_{i}$ given $\mathcal{S}$ if and only if for every $%
\overline{S}_{-i}\in \mathcal{S}$, there exists $\overline{s}_{i}\in S_{i}(h,%
\overline{a}_{i})$ such that%
\begin{equation*}
\min_{s_{-i}\in \overline{S}_{-i}}u_{i}\left( \zeta (\overline{s}%
_{i},s_{-i})\right) \geq \max_{z\in Z(h,a_{i},\overline{S}_{-i})}u_{i}\left(
z\right) .
\end{equation*}
\end{remark}

In the case of perfect contingent reasoning, s-dominance does not even
require to come up with a continuation strategy after the dominating action:
tailoring the continuation strategy to each contingency is equivalent to
just identifying the best possible payoff given each contingency. The
following definition formalizes this special case.

\begin{definition}
Action $\overline{a}_{i}\in A_{i}^{h}$ wishfully dominates action $a_{i}\in
A_{i}^{h}$ if%
\begin{equation}
\forall s_{-i}\in S_{-i}(h),\text{ \ \ }\max_{z\in Z(h,\overline{a}%
_{i},s_{-i})}u_{i}\left( z\right) \geq \max_{z\in
Z(h,a_{i},s_{-i})}u_{i}\left( z\right) ,  \label{WI1}
\end{equation}%
Action $\overline{a}_{i}$ is wishfully dominant if it wishfully dominates
every other $a_{i}\in A_{i}^{h}$.
\end{definition}

\begin{remark}
Action $\overline{a}_{i}$ wishfully dominates $a_{i}$ if and only if $%
\overline{a}_{i}$ s-dominates $a_{i}$ given the singleton partition $%
\mathcal{S}=\left\{ \left\{ s_{-i}\right\} |s_{-i}\in S_{-i}(h)\right\} $.
\end{remark}

We will illustrate wishful dominance by example in our dynamic TTC\
mechanism of Section \ref{Section: TTC}. The term \textquotedblleft
wishful\textquotedblright\ is justified by the fact that the player looks at
the best payoffs she could obtain under each contingency, although,
typically, no single continuation strategy can achieve the best outcome in
all contingencies. This observation also highlights a fundamental
inconsistency between planning and wishful dominance:\ a player cannot, at
the same time, be a planner and reason according to wishful dominance.%
\footnote{%
Games with perfect and complete information are an exception, in that
looking for the best outcome in each contingency pins down a well-defined
plan. We conjecture that in this class of games, wishful dominance, weak
dominance, and even obvious dominance coincide.}

Wishful dominance allows us to provide a characterization of
strategy-proofness that does not involve continuation strategies. By theorem %
\ref{CHAR 1}, strategy-proofness is equivalent to the existence of a weakly
dominant action at every information set that is consistent with the
dominant strategy. Thus, by Theorem \ref{Theorem local dominant = local
c-dominant}, the same equivalence holds with wishful in place of weak
dominance.

\begin{theorem}
\label{Corollary: lwd=wd}A game is strategy-proof if and only if, for each
player $i$, there exists a strategy $\overline{s}_{i}$ such that $\overline{s%
}_{i}(h)$ is wishfully dominant at every $h\in H_{i}^{\ast }(\overline{s}%
_{i})$.
\end{theorem}

Theorem \ref{Corollary: lwd=wd} says that the search for a weakly dominant
strategy can be decomposed into local problems that do not even require to
conceive continuation strategies. In other words, strategy-proofness is
robust to the inability to plan forward, if players analyze each contingency
in isolation. The possibility to discover the dominant actions without even
entertaining continuation strategies strengthens our argument for the
simplicity of strategy-proof dynamic mechanisms --- in our dynamic TTC\
mechanism, searching for wishfully dominant actions will be particularly
easy.

\section{Local Dominance\label{Section: very local dominance}}

The notions of dominance we introduced so far are silent as to whether the
two actions can be ranked using a partition and continuation strategies that
are truly easy to identify. In this section, we endogenize the partition of
the contingencies and restrict the use of continuation strategies according
to the simplicity principles outlined in the introduction. While doing so,
as in s-dominance, we maintain the idea that players consider each scenario
separately. This is a natural choice given that, with local dominance, we
aim to capture the separate treatment of present and future of the game in
the player's mind.

\paragraph{Endogenizing the partition}

Fix an information set $h\in H_{i}$. For each action $a_{i}\in A_{i}^{h}$,
let $S_{-i}^{a_{i},h}$ denote the set of contingencies in which the game
will end for player $i$ after choosing $a_{i}$ at $h$, i.e., there is no
further information set where $i$ is active after playing $a_{i}$ at $h$:%
\begin{equation*}
S_{-i}^{a_{i},h}=\left\{ s_{-i}\in S_{-i}(h):\forall z\in Z(h,a_{i},s_{-i}),%
\NEG{\exists}h^{\prime }\in H_{i}^{\ast },h\prec h^{\prime }\prec z\right\} .
\end{equation*}%
Given an action pair $(\overline{a}_{i},a_{i})\in A_{i}^{h}\times A_{i}^{h}$%
, let%
\begin{equation*}
\begin{array}{lll}
S^{\overline{a}_{i},a_{i}}(h)=S_{-i}^{a_{i},h}\cap S_{-i}^{\overline{a}%
_{i},h} & \text{ \ \ } & S_{a_{i}}^{\overline{a}_{i}}(h)=S_{-i}^{\overline{a}%
_{i},h}\backslash S_{-i}^{a_{i},h} \\ 
S_{\overline{a}_{i}}^{a_{i}}(h)=S_{-i}^{a_{i},h}\backslash S_{-i}^{\overline{%
a}_{i},h} & \text{ \ \ } & S_{\overline{a}_{i},a_{i}}(h)=S_{-i}(h)\backslash
(S_{-i}^{\overline{a}_{i},h}\cup S_{-i}^{\overline{a}_{i},h}).%
\end{array}%
\end{equation*}%
In words, each of these sets contains the strategies of the opponents that
terminate our player's game if she chooses the action(s) at the superscript,
and do not terminate it if she chooses the action(s) at the subscript. Let 
\begin{equation*}
\mathcal{S}^{\ell }(h,\overline{a}_{i},a_{i})=\left\{ S^{\overline{a}%
_{i},a_{i}}(h),S_{\overline{a}_{i}}^{a_{i}}(h),S_{a_{i}}^{\overline{a}%
_{i}}(h),S_{\overline{a}_{i},a_{i}}(h)\right\} .
\end{equation*}%
We call $\mathcal{S}^{\ell }(h,\overline{a}_{i},a_{i})$ the \textbf{local
partition}.

The local partition seems natural from a local viewpoint, for various
reasons. For a player who compares the current actions with a blurred view
of the future, it is natural to first focus on their possible immediate
consequences. This requires the player to understand in what circumstances
each action will be the last action she plays and hence will directly yield
the final outcome. This way of partitioning is indeed \textquotedblleft
local\textquotedblright , in that the scenarios only depend on the moves of
the opponents before our player's next active stage, and can be identified
as long as she can conceive her next moves. Formally, the local partition is
measurable\ with respect to the information our player receives at the next
decision nodes. For each $\widetilde{a}_{i}=\overline{a}_{i},a_{i}$, call $%
\mathcal{H}^{\ast }(h,\widetilde{a}_{i})$ the set of the first \emph{active}
information sets of $i$ after choosing $\widetilde{a}_{i}$ at $h$, and let $%
\mathcal{S}^{\ast }(h,\widetilde{a}_{i})$ denote the collection $\left\{
S_{-i}(h^{\prime })|h^{\prime }\in \mathcal{H}^{\ast }(h,\widetilde{a}%
_{i})\right\} \cup \{S_{-i}^{\widetilde{a}_{i},h}\}$. The local partition, $%
\mathcal{S}^{\ell }(h,\overline{a}_{i},a_{i})$, is weakly coarser than the
meet of the partitions $\mathcal{S}^{\ast }(h,\overline{a}_{i})$ and $%
\mathcal{S}^{\ast }(h,a_{i})$.

\paragraph{Mimicking strategies}

Next, we formalize the idea of comparing two actions \textquotedblleft
ceteris paribus\textquotedblright\ with respect to the future moves, that
is, under the hypothesis of \textquotedblleft continuing in the same
way\textquotedblright\ after the two. Coherently with analysing each
scenario in isolation, we are going to talk of continuing in the same way
conditional on the particular scenario under consideration.

Let $\tau (\widetilde{h})$ denote the stage of an information set $%
\widetilde{h}$. Fix $\overline{s}_{i},s_{i}\in S_{i}(h)$ and a scenario $%
\overline{S}_{-i}\subseteq S_{-i}(h)$. We say that $\overline{s}_{i}$ 
\textbf{mimics }$s_{i}$ given $h$ and $\overline{S}_{-i}$ when, for every $%
s_{-i}\in \overline{S}_{-i}$, the following condition holds: For every $%
\overline{h}\in \mathcal{H}^{\ast }(h,\overline{s}_{i}(h))$ and for every $%
\overline{h}^{\prime },h^{\prime }\in H_{i}$ such that $\overline{h}\preceq 
\overline{h}^{\prime }\prec \zeta (\overline{s}_{i},s_{-i})$, $h^{\prime
}\prec \zeta (s_{i},s_{-i})$, and $\tau (\overline{h}^{\prime })=\tau
(h^{\prime })$, we have $\overline{s}_{i}(\overline{h}^{\prime
})=s_{i}(h^{\prime })$. In words, under every element of $\overline{S}_{-i}$%
, once player $i$ becomes active again after choosing $\overline{s}_{i}(h)$
at $h$, the mimicking strategy prescribes the same action as the mimicked
strategy at each subsequent stage. Note that mimicking only starts at the
next active information set because, before that, a player is forced to play
a dummy action.

Mimicking is a strong requirement because it is formulated \textquotedblleft
ex-post\textquotedblright : for each realization of $s_{-i}$ in $\overline{S}%
_{-i}$, the actual sequence of moves of player $i$ (from the first active
stage after $\overline{a}_{i}$) must be the same under $s_{i}$ and its
mimicker $\overline{s}_{i}$.

\paragraph{Irrelevance}

Now we introduce our version of \textquotedblleft sure thing
principle\textquotedblright\ for the problem at hand, that is, the idea that
a scenario is \textquotedblleft irrelevant\textquotedblright\ for the
comparison of two actions. Our player will conclude that a scenario is
irrelevant when, by continuing in the same way\ after the two actions, she
would obtain the same outcome. This observation is different in nature from
the comparison of best and worst outcomes we considered so far. Typically,
it follows from a symmetric structure of the game that is easy to spot
without actual scrutiny of all the possible ways she and the opponents may
continue playing, and of the consequent outcomes.

Formally, we say that a scenario $\overline{S}_{-i}$ is \textbf{irrelevant }%
for $(\overline{a}_{i},a_{i})$ at $h$ when, for every $s_{i}\in
S_{i}(h,a_{i})$, there exists $\overline{s}_{i}\in S_{i}(h,\overline{a})$
that mimics $s_{i}$ given $h$ and $\overline{S}_{-i}$ such that%
\begin{equation}
\forall s_{-i}\in \overline{S}_{-i},\text{ \ \ \ }g_{i}\left( \zeta (%
\overline{s}_{i},s_{-i})\right) =g_{i}\left( \zeta (s_{i},s_{-i})\right) .
\label{IRRELEVANCE}
\end{equation}%
We define irrelevance for an \emph{ordered} pair of actions because we have
in mind a player who has already concluded that $\overline{a}_{i}$ is better
than $a_{i}$ in terms of possible immediate consequences. To check that this
ranking is not overturned when looking at the possible future consequences,
it is enough to conclude that for any outcome that she can obtain after $%
a_{i}$, she can also obtain it after $\overline{a}_{i}$.

Ignoring an irrelevant scenario is an application of the sure thing
principle in which the irrelevant states are obtained under the hypothesis
of continuing in the same way. The sure thing principle is sometimes
violated in other contexts, and one of the reasons could be that, in the
problem at hand, the irrelevant states are naturally pooled with relevant
ones into a non-irrelevant scenario in the decision-maker's mind. For
instance, in Ellsberg's problem with 30 red balls and 60 yellow or blue
balls, the state \textquotedblleft blue\textquotedblright\ is irrelevant for
the comparison of \textquotedblleft bet on red\textquotedblright\ and
\textquotedblleft bet on yellow\textquotedblright , but it is naturally
pooled with the other losing state \textquotedblleft
yellow\textquotedblright\ under the first bet and with the other losing
state \textquotedblleft red\textquotedblright\ under the second bet. In our
context, we expect a player to identify the scenario \textquotedblleft the
game continues after both actions\textquotedblright\ before she recognizes
its irrelevance, for the reasons we outlined before, which have nothing to
do with the outcome function.

\paragraph{Local dominance}

We are now ready to formalize the idea of comparing actions under the local
partition and the hypothesis of \textquotedblleft continuing in the same
way\textquotedblright .

\begin{definition}
\label{Def: local dominance}Fix an information set $h\in H_{i}^{\ast }$ and
an action pair $\left( \overline{a}_{i},a_{i}\right) \in A_{i}^{h}\times
A_{i}^{h}$. Action $\overline{a}_{i}$ \textbf{locally dominate}s action $%
a_{i}$ if for each non-empty $\overline{S}_{-i}\in \mathcal{S}^{\ell }(h,%
\overline{a}_{i},a_{i})$, for each $s_{i}\in S_{i}(h,a_{i})$, there exists $%
\overline{s}_{i}\in S_{i}(h,\overline{a}_{i})$ that mimics $s_{i}$ given $h$
and $\overline{S}_{-i}$ such that:%
\begin{equation}
\begin{array}{cc}
\text{if }\overline{S}_{-i}\not=S_{\overline{a}_{i},a_{i}}(h)\text{,} & 
\min\limits_{s_{-i}\in \overline{S}_{-i}}u_{i}\left( \zeta (\overline{s}%
_{i},s_{-i})\right) \geq \max\limits_{s_{-i}\in \overline{S}%
_{-i}}u_{i}\left( \zeta (s_{i},s_{-i})\right) ; \\ 
\text{if }\overline{S}_{-i}=S_{\overline{a}_{i},a_{i}}(h)\text{,} & 
g_{i}\left( \zeta (\overline{s}_{i},\cdot )\right) |_{\overline{S}%
_{-i}}=g_{i}\left( \zeta (s_{i},\cdot )\right) |_{\overline{S}_{-i}}\text{ }
\\ 
& \text{(i.e., }S_{\overline{a}_{i},a_{i}}(h)\text{ is irrelevant).}%
\end{array}
\label{Eq: local D}
\end{equation}

Action $\overline{a}_{i}$ is locally dominant if it locally dominates every
other $a_{i}\in A_{i}^{h}$.
\end{definition}

Local dominance builds on s-dominance by considering each scenario of the
local partition in isolation. While s-dominance assumes that, in each
scenario, a player can always find a good-enough continuation strategy after
the dominating action, local dominance does not assume this ability; it only
requires to entertain the salient continuation strategy that
\textquotedblleft continues in the same way\textquotedblright\ as after the
dominated action, i.e., the mimicking strategy. In the way outcomes are
compared, local dominance departs from (s-)dominance in the scenario where
the game continues after both actions: instead of assuming that a player can
find worst and best outcomes after the two actions, it assumes that a player
can realize that the current choice has no impact on the continuation game.
Of course, a cognitively limited player could also ignore this scenario
simply because it seems too complicated --- local dominance is robust to
these considerations.

Local dominance can be interpreted as a new kind of one-shot deviation
check. In particular, local dominance evaluates a switch from the dominating
to the dominated action not under a fixed continuation strategy, which may
prescribe different future moves after the two actions, but under the idea
of \textquotedblleft continuing in the same way\textquotedblright\ after the
two actions, no matter how. In this sense, local dominance is a game
theoretical translation of the \textquotedblleft ceteris
paribus\textquotedblright\ principle. For the scenario in which the game
continues after both actions, the invariance of final outcomes typically
requires the continuation play of the opponents to be the same as well. In
turn, this typically requires that the opponents do not observe (and hence
cannot condition their choices on) whether our player chose one action or
the other. Indeed, like weak dominance and unlike obvious dominance, local
dominance is sensitive to how informed the opponents are of a player's
moves: roughly speaking, the more they observe, the more the opportunities
to strategize, the more difficult for the player to identify a dominant
choice.

A peculiarity of local dominance is not being transitive. The reason is that
the local partition changes with the action pair under comparison. We do not
find this intransitivity undesirable for a notion of dominance that aims to
capture an idea of simplicity:\ \textquotedblleft similar\textquotedblright\
enough alternatives may be easy to rank, too \textquotedblleft
dissimilar\textquotedblright\ ones may not.

Definition \ref{Def: local dominance} formalizes the idea of comparing
actions under the view of continuing in the same way. In particular, it
matches each continuation strategy after the dominated action with a
mimicking continuation strategy after the dominating action, mirroring
Definition \ref{Definition S-dominance} of s-dominance. But just like
s-dominance, a player can assess local dominance without scrutinizing the
continuation strategies after the dominated action (see Remark \ref{Remark:
simplification local c-dom}); moreover, under the local partition, in the
scenarios where the dominating action terminates the game, there is clearly
no need to entertain any continuation strategy after the dominating action
either. Thus, by the very nature of the local partition, our player can
analyze each scenario as follows.

\begin{remark}
\label{Rem: local dominance char}Action $\overline{a}_{i}$ locally dominates
action $a_{i}$ if and only if the following conditions hold (when the
respective scenario is non-empty):

\begin{enumerate}
\item (comparison of immediate consequences)%
\begin{equation}
\min_{z\in Z(h,\overline{a}_{i},S^{\overline{a}_{i},a_{i}}(h))}u_{i}\left(
z\right) \geq \max_{z\in Z(h,a_{i},S^{\overline{a}_{i},a_{i}}(h))}u_{i}%
\left( z\right) ;  \label{VLDbis}
\end{equation}

\item 
\begin{equation}
\min_{z\in Z(h,\overline{a}_{i},S_{a_{i}}^{\overline{a}_{i}}(h))}u_{i}\left(
z\right) \geq \max_{z\in Z(h,a_{i},S_{a_{i}}^{\overline{a}%
_{i}}(h))}u_{i}\left( z\right) ;  \label{VLDbisbis}
\end{equation}

\item there exists $\overline{s}_{i}\in S_{i}(h,\overline{a}_{i})$ that
mimics any $s_{i}\in S_{i}(h,a_{i})$ given $h$ and $S_{\overline{a}%
_{i}}^{a_{i}}(h)$ such that%
\begin{equation}
\min_{s_{-i}\in S_{\overline{a}_{i}}^{a_{i}}(h)}u_{i}\left( \zeta (\overline{%
s}_{i},s_{-i})\right) \geq \max_{z\in Z(h,a_{i},S_{\overline{a}%
_{i}}^{a_{i}}(h))}u_{i}\left( z\right) ;  \label{VLDtris}
\end{equation}

\item $S_{\overline{a}_{i},a_{i}}(h)$ is irrelevant.
\end{enumerate}
\end{remark}

To see the equivalence between conditions (\ref{VLDbis}),(\ref{VLDbisbis})
and condition (\ref{Eq: local D}), note that in scenarios $S^{\overline{a}%
_{i},a_{i}}(h)$ and $S_{a_{i}}^{\overline{a}_{i}}(h)$ there is no active
information set of player $i$ after $\overline{a}_{i}$. Therefore, the only
\textquotedblleft continuation strategy\textquotedblright\ after $\overline{a%
}_{i}$ (the one that prescribes the dummy actions until stage $T$) trivially
mimics any continuation strategy after $a_{i}$ --- recall that mimicking is
only required to start from the next active information sets onwards. As for
scenario $S_{\overline{a}_{i}}^{a_{i}}(h)$, there is just one mimicking
strategy to consider, because, within this scenario, all $s_{i}\in
S_{i}(h,a_{i})$ induce the same dummy continuation strategy after $a_{i}$.

We say that a game is \textbf{locally strategy-proof} when each player has a
locally dominant action at every information set that can be reached if she
always chooses her locally dominant actions. In Section 5.1, we show that
the ascending auctions we considered in the introduction are locally
strategy-proof, and we construct a locally strategy-proof mechanism that
implements the TTC allocation rule.

\paragraph{Comparison with other notions of dominance between actions}

We start by comparing local dominance with the weakest notion of dominance
we introduced, wishful dominance. Suppose that action $\overline{a}_{i}$
locally dominates action $a_{i}$; then, under each contingency $s_{-i}$,
compared to the best outcome after $a_{i}$, after $\overline{a}_{i}$ one can
achieve an outcome that is at least as good (in particular, the $\emph{same}$
outcome if $s_{-i}\in S_{\overline{a}_{i},a_{i}}(h)$). Therefore, local
dominance refines wishful dominance. With this, Theorem \ref{Corollary:
lwd=wd} implies that local\ strategy-proofness refines strategy-proofness.

\begin{proposition}
\label{Prop: local-wishful}If $\overline{a}_{i}$ locally dominates $a_{i}$,
then it wishfully dominates $a_{i}$. Therefore, if a game is locally
strategy-proof, it is strategy-proof.
\end{proposition}

Local dominance is neither weaker nor stronger than s-dominance if we fix
the local partition. The reason why it is not stronger, despite the
restrictions on the use of continuation strategies, is that, as we observed,
irrelevance is just a different criterion\ from the comparison of worst and
best payoffs.

Similarly, local dominance is not weaker or stronger than obvious dominance
between actions. In static games, the two notions actually coincide, because
the only non-empty scenario is the one in which the game ends after every
action. In our framework, a game is static if $T=1$, so there is stage $0$
for the move of nature and only stage $1$ for players' moves.

\begin{proposition}
\label{Prop: e-dom = obv-dom}Fix a static game, an information set $h\in
H_{i}^{\ast }$, and an action pair $\left( \overline{a}_{i},a_{i}\right) \in
A_{i}^{h}\times A_{i}^{h}$. Then, action $\overline{a}_{i}$ locally
dominates action $a_{i}$ if and only if $\overline{a}_{i}$ obviously
dominates $a_{i}$.
\end{proposition}

Proposition \ref{Prop: e-dom = obv-dom} allows to establish a connection
between local and obvious strategy-proofness, which we discuss in Section %
\ref{Section: comparison with OD}.

\section{Applications}

\subsection{Ascending-price auctions}

We analyze with local dominance the ascending-price auction we outlined in
the introduction, and its variant with the waiting option. The one without
the waiting option is the auction format that was shown easy to play by the
experiment of Kagel et al. (1987). There are $n$ bidders with private
valuations for the object, which for simplicity we assume to be integer.%
\footnote{%
With non-integer valuations, the valuation of a bidder may be higher than
the current price but lower than the next price, and then bidding would not
locally dominate leaving. Nonetheless, we could still claim that players
will bid \emph{at least} up to that price.} At each stage $t=1,...,T$ (with $%
T$ larger than any possible valuation), player only observe whether the
object was already assigned or not, and if not, the players who have not
left before simultaneously choose between bid ($b$) and leave ($\ell $); in
presence of the additional waiting option, they can also wait ($w$).
Formally, the players who left before stage $t$ are forced to play $\ell $,
and all players must play $\ell $ once the object is assigned. The
\textquotedblleft stage-$t$ outcome rule\textquotedblright\ is the
following: if no player bids, the object is assigned at random at price $p=t$
among the bidders who have not left before stage $t$; if only one player
bids, she wins the object at price $p=t$; if more than one player bids, the
auction moves to round $t+1$. If waiting is allowed, the stage-$t$ outcome
rule makes no distinction between waiting and leaving. Thus, the only
difference between waiting and leaving is that, if the auction continues, a
player who waited can move at the next stages, while a player who left must
play $\ell $ forever.

\begin{proposition}
\label{Prop: auction}The ascending-price auction is locally strategy-proof
(also with the waiting option). In particular, for every player, it is
locally dominant to bid as long as the price is below her valuation and then
leave when the price reaches her valuation.
\end{proposition}

Here we provide a sketch of the proof, which is formalized in the Appendix.

Take first the viewpoint of a player at a stage where the price is still
below her valuation. What makes the decision to bid easy to take? We start
from the comparison with leaving. It is probably natural for a player to
first focus on the scenario in which bidding terminates the game and
immediately yields the final outcome. This occurs when no other player bids.
In this scenario, the outcome of bidding is winning the object, the outcome
of leaving is the lottery. Thus, in terms of possible immediate
consequences, bidding is better than leaving. Then, our player also realizes
that if she bids and the game continues, her outcome will be determined in
the future. To restrict the realm of the possible future outcome after
bidding, it is probably natural to entertain the idea of leaving at the next
stage. With this continuation strategy, our player cannot incur a loss after
bidding at the current stage, therefore the comparison of local outcomes is
not overturned. So, bidding locally dominates leaving.

Now we move to the comparison between bidding and waiting. In the scenario
where bidding terminates the game (i.e., no one else bids), waiting is
equivalent to leaving, and thus bidding is better than waiting. When bidding
does not terminate the game, waiting may or may not terminate the game. When
only one opponent bids, waiting terminates the game. Then, also for the
comparison between bidding and waiting, it is probably natural to entertain
the idea of leaving (and thus terminating the game)\ at the next stage. So,
for the same argument of the comparison with leaving, bidding is better than
waiting. When more than one opponent bids, the game continues also after
waiting. In this scenario, it is easy to recognize that the choice between
bidding and leaving has no impact whatsover on the final outcome. In greater
detail, bidding or waiting does not alter what the opponents observe, and
hence what they will do, nor what our player will observe and be allowed to
do. Note also that the outcome of a stage only depends on players' moves at
that stage. Thus, the scenario in which the auction continues after both
bidding and leaving is irrelevant for the comparison. Hence, bidding locally
dominates waiting.

Take now the viewpoint of a bidder at the stage where the price reaches her
valuation. What makes the decision to leave easy to take? If the auction
ends, not matter our player's action, her surplus will be zero. Leaving can
only have this immediate consequence, while bidding and waiting can also
entail that our player's outcome will be determined in the future. However,
no matter what this outcome will be, it cannot bring a positive payoff to
our player. Thus, leaving locally dominates bidding and, if available,
waiting.

\subsection{Top Trading Cycles\label{Section: TTC}}

Consider the classical object allocation problem without monetary transfers.
There are $n$ agents and $N$ objects. Each agent initially owns an object
and has a strict preference ranking over all objects. A rule specifies for
every preference profile a reallocation of the objects to the agents such
that each agent gets exactly one object. A prominent rule that has been
extensively studied in the literature is the TTC rule. It adopts an
iterative algorithm proposed by Gale. At every iteration, the algorithm
generates a directed graph in which the nodes are the agents who are yet to
be assigned an object and the arrows link each agent to the owner of her
highest-ranked object that is still available. Such a graph always has at
least one cycle, and the agents that end up in a cycle are assigned the
object they are pointing to.

The TTC rule is known to be strategy-proof: in the direct mechanism,
reporting the true preferences is weakly dominant. However, the direct
mechanism is known to be difficult to play in practice. In particular, a
player might be tempted to rank an object $b$ above an object $a$ despite
preferring $a$ to $b$, because she fears that she might miss her chance to
get object $b$ while the algorithm keeps pointing her to object $a$
unsuccessfully. To address these concerns, we translate Gale's TTC algorithm
into a dynamic mechanism with three simplicity features. First, at each
stage, players are only asked to name one of the still-available objects. In
this way, we decompose the problem of revealing your own preferences into a
sequence of smaller partial-revelation problems. Second, players cannot move
until the last object they named is assigned to someone else. This reassures
players that whenever an opportunity for trade pops up (i.e., some other
player points to her directly or indirectly),\footnote{%
We say that player $j$ points to the object of player $i$ indirectly if
there is a sequence of players starting with $j$ and ending with $i$ such
that each player in the sequence names the object owned by the next player.}
it remains intact through time and can be exploited later. Third, our
mechanism carefully releases information to players so to reassure them
that, if the game continues after both $a$ and $b$, they can continue in the
same way, and then will also get the same outcome. To allow players to
continue in the same way, we let them observe the set of available outcomes
also when they do not move, so that their available information does not
depend on how long they have to wait.\footnote{%
Formally, as we give the traditional representation of information flows
just at the information sets where players are active, they observe the past
history of available objects once they get to move again. The observability
of such history, and not just of the current menu of available objects,
distinguishes our mechanism from that of Bo and Hakimov (2022).} To
guarantee the same final outcome, we do not reveal players' choices to the
opponents, so that one does not have to worry that her choices may affect
the opponents' future choices, in a way that negatively affects her final
outcome. With this, we show that in our mechanism naming the favorite object
is always locally dominant, whereas in the direct mechanism submitting the
true preference ranking is not locally dominant (i.e., by Proposition \ref%
{Prop: obvious=local}, obviously dominant). Therefore, we obtain the
following positive result, which contrasts with Li's (2017) impossibility of
obvious strategy-proof implementation of the TTC\ rule.

\begin{theorem}
\label{Theorem: TTC}The TTC\ rule can be implemented with a locally
strategy-proof mechanism.
\end{theorem}

To prove the theorem, we explicitly construct such a mechanism, which we
call \textit{dynamic TTC\ mechanism}. At stage $1$, every player names one
object. The players who end up in a cycle are assigned the object they
named. At each stage $t=2,...,N$, \emph{players only observe which objects
are still available}, i.e., which objects have not been assigned in the
previous stages. If the last object a player named is still available, then
the player cannot modify her choice --- to comply with our formalism, she is
obliged to play the dummy action of renaming the same object. Analogously, a
player who has already been assigned an object keeps naming that object. If
instead the last object a player named becomes unavailable at the current
stage, then she must name one of the available objects. Again, the players
who end up in a cycle trade their objects. Since at least one cycle occurs
at every stage, by stage $N$ the assignment is complete and players leave
with their assigned object.

In the dynamic TTC mechanism, naming your favorite available object is
locally dominant at every information set. In the next paragraph we provide
a skectch of the proof, which we formalize in the Appendix.

Take the point of view of a player at an information set where she must name
a new object. Let $a$ be her favorite available object and $b$ another
available object. Is naming $a$ better than naming $b$? If naming $a$
immediately yields $a$ --- without a doubt. Now suppose that after naming $a$
our player must move again at some stage $t$, because $a$ was assigned to
someone else. Then, she may wonder whether she will still be in time to
catch up with $b$ or any other object she would name after $b$ at stage $t$,
and then continue in the same way, so to obtain the same object (the object
one gets is always the final object one names). The answer is yes: whatever
object our player would be naming at stage $t$ after $b$ will still be
available after $a$. As long as our player is not assigned an object, her
moves do not affect the set of available objects, and thus do not affect the
moves of the opponents. Therefore, if after naming $b$ our player is not
assigned an object before stage $t$, she will face the same set of available
objects at stage $t$ after naming $a$ and after naming $b$. If instead after 
$b$ our player ends up in a cycle and is assigned an object at a stage $%
t^{\ast }<t$, that object must remain available until stage $t$ also after $%
a $, because the other members of the cycle cannot modify their choices
between stages $t^{\ast }$ and $t$.

Note that our dynamic TTC mechanism actually requires a simpler form of
contingent reasoning than the local partition. The above argument only
requires our player to distinguish the scenario in which the locally
dominant action $a$ terminates the game from the scenario in which it does
not, and deem the latter scenario irrelevant, i.e., she can get the same
object by continuing in the same way. This bipartition is entirely driven by
the dominant action, which is very easy to recognize as the right candidate
choice\ based on the comparison of the possible immediate consequences.
Then, our player can use the same (bi)partition to finalize the comparison
with all the alternatives.

If players are capable of perfect contingent reasoning, they can also find
their optimal actions by reasoning according to wishful dominance. In our
dynamic TTC game, player $i$ can ask herself: \textquotedblleft Suppose that
after playing $b$ the best object I can obtain is $c$. Can I\ get something
at least as good after playing $a$?\textquotedblright\ The answer is yes. If
player $i$ does not get $a$ (which is the best she can get), she will have
the opportunity of naming $c$ next, and then she will actually get $c$. The
reason is that, as long as player $i$ does not get an object, she does not
affect what the opponents observe, and hence she does not affect their
moves. So, in a contingency where she can get $c$ after naming $b$, all the
opponents in the cycle that gives her $c$ make the same moves also when she
names $a$ in place of $b$, so she can close that cycle by naming $c$ after
naming $a$, if she does not get $a$.

Proving that naming the favorite object is wishfully dominant is somewhat
simpler than proving that it is locally dominant, because it does not
require to check that the object $c$ obtained after naming $b$ can also be
obtained after naming $a$ \emph{by continuing in the same way}. That proving
local dominance requires more work is natural, since it is a stronger notion
than wishful dominance (see Proposition \ref{Prop: local-wishful});
nonetheless, the two notions capture the same intuition that we expect
players to have in this game: there is nothing to lose from naming the
favorite object. In other games, there could be wishfully dominant actions
that are not locally dominant. In such a case, local dominance, which is a
stronger simplicity standard, deems the intuitions captured by wishful
dominance too difficult for real players.

\section{Comparison with the literature}

\subsection{Local dominance versus obvious dominance\label{Section:
comparison with OD}}

In static games, local strategy-proofness and obvious strategy-proofness
coincide. This is a consequence of Proposition \ref{Prop: e-dom = obv-dom}
and Theorem \ref{CHAR 1}: by Proposition \ref{Prop: e-dom = obv-dom}, at an
information set of a static game (i.e., given a \textquotedblleft
type\textquotedblright ), an action is locally dominant if and only if it is
obviously dominant, and hence, by Theorem \ref{CHAR 1}, a strategy is
obviously dominant if and only if it prescribes an obviously dominant action
at every information set.

\begin{proposition}
\label{Prop: obvious=local}A static game is obviously strategy-proof if and
only if it is locally strategy-proof.
\end{proposition}

Thus, like obvious dominance, local dominance rules out most direct
mechanisms as too complicated.

In dynamic games, in one dimension, local dominance adopts a stricter
simplicity standard than obvious dominance, in that it imposes the use of
mimicking continuation strategies, rather than admitting the use of the
optimal continuation strategy. In the dimension of contingent reasoning,
instead, local dominance is more permissive than obvious dominance, in that
it introduces the local partition. For this reason, local dominance is not
stronger than obvious dominance, as shown by our dynamic TTC mechanism.
Similarly, because of the complete lack of contingent reasoning, obvious
dominance cannot explain why the ascending-price auction with simultaneous
moves was found easy to play by Kagel et al. (1987), as we show next.

Differently from local dominance, obvious dominance assumes that the bidder
can formulate at the outset the plan of bidding until the price reaches her
valuation. Yet, according to obvious dominance, she cannot establish the
superiority of this \textquotedblleft sincere strategy\textquotedblright\
over a \textquotedblleft stingy strategy\textquotedblright\ that leaves at a
lower price $p$. This is because, when she compares the two strategies, she
considers at the same time the chance of winning the object with the stingy
strategy (when all the opponents leave at price $p$), and the possibility of
not winning the object with the sincere strategy (when some opponent bids up
to a higher price than our bidder's valuation). According to local
dominance, instead, the chance of winning after leaving is liquidated in the
primary\ scenario where the auction ends also after bidding, and in this
scenario bidding is clearly better than leaving, as it guarantees winning
the object. The experimental findings of Kagel et al. (1987) suggest that
players do not mix up this \textquotedblleft local\textquotedblright\
scenario with the alternative scenario in which bidding does not terminate
the auction.

\subsection{Comparison with Pycia and Troyan (2022)}

The paper that is closest to ours is Pycia and Troyan (2022), who introduce
the notion of \emph{simple dominance}. At each decision node, players can
only plan for a given set of \textquotedblleft simple\textquotedblright\
future nodes and consider a \textquotedblleft strategic
plan\textquotedblright\ for those nodes. A strategic plan is simply dominant
when the worst outcome that is consistent with it is not worse than the best
outcome that the player may obtain after any alternative action at the
current node. Thus, compared to obvious dominance, simple dominance
considers the larger set of outcomes that are consistent not only with all
the possible future moves of the opponents, but also with all possible own
moves at the non-simple future nodes.

Among our notions of dominance, the closest to simple dominance is obvious
dominance between actions, because, like simple dominance, it does not rely
on any form of contingent reasoning. To facilitate the comparison, we
provide the following characterization of our notion.

\begin{remark}
\label{Remark: char obv dom}Fix an information set $h\in H_{i}^{\ast }$.
Action $\overline{a}_{i}\in A_{i}^{h}$ is obviously dominant if and only if,
for every $a_{i}\in A_{i}^{h}\backslash \left\{ \overline{a}_{i}\right\} $,
there exists $\overline{s}_{i}\in S_{i}(h,\overline{a}_{i})$ such that%
\begin{equation*}
\min_{s_{-i}\in S_{-i}(h)}u_{i}\left( \zeta (\overline{s}_{i},s_{-i})\right)
\geq \max_{z\in Z(h,a_{i},\overline{S}_{-i})}u_{i}\left( z\right) .
\end{equation*}
\end{remark}

Remark \ref{Remark: char obv dom} is a corollary of Remark \ref{Remark:
simplification local c-dom}, as obvious dominance between actions coincides
with s-dominance under the trivial partition.

The difference between the notion of simply dominant strategic plan and the
notion of obviously dominant action is that the first specifies not just the
current action but also the actions at the future decision nodes the player
can plan for. This (partial)\ continuation plan has to beat all the
alternatives at the current node. Instead, an obviously dominant action is
not associated with one \emph{fixed} continuation strategy: as shown by
Remark \ref{Remark: char obv dom},\ our player can entertain different
continuation strategies for the comparison with different alternatives.\ But
even if we allow for flexibility in the use of continuation strategies, the
issue with both notions is that finding the optimal (partial) continuation
strategy may be the only way to establish dominance.

Pycia and Troyan (2022) eradicate this problem by focusing on \emph{strong
obvious dominance}, the special case of simple dominance in which the player
does not perceive any future decision as simple, and thus does not plan at
all. Thus, an action $\overline{a}_{i}$ is strongly obviously dominant when
the worst outcome that follows it is not worse than the best outcome that
follows any alternative action. Such outcomes are computed across all the
possible future moves of the player herself, because she has no clue of how
she herself will play.

With local dominance, we let our player compare first the possible outcomes
in case the game immediately ends, which do not depend on the future moves.
This separation between present and future outcomes is a coarse form of
contingent reasoning. When it comes to comparing the possible future
outcomes, we introduce some simple considerations about the continuation
game that ease the task. For the scenario in which the game continues after
the candidate dominating action $a$ but not after the alternative action $b$%
, our player simplifies her view of the possible future outcomes after $a$
with a salient \textquotedblleft mimicking strategy\textquotedblright , such
as terminating the game at the next stage, or reverting to $b$. While any
continuation strategy is a legitimate way of restricting the realm of
possible outcomes after the candidate dominating action, by using the
mimicking strategy we endogenize which continuation strategy is simple to
conceive for the player, without requiring any consideration on the
optimality of future moves. For the most complicated scenario in which the
game continues after both actions, our player only checks whether the
current choice may matter at all for the final outcome --- the notion of
irrelevance. Overall, compared to strong obvious dominance, we also consider
a player who has no clue of how she will play in the future, but we
radically differ in the way the player tackles this problem.

Because of irrelevance, local dominance is not weaker than strong obvious
dominance. Indeed, irrelevance is just a different, not a weaker criterion
than the comparison of the best and the worst possible future outcomes. The
following example illustrates this point.\footnote{%
This image is borrowed from https://www.youtube.com/watch?v=yuf8cd1eypA.}

\FRAME{dhF}{3.1324in}{1.7659in}{0pt}{}{}{mmexport1663645496832.png}{\special%
{language "Scientific Word";type "GRAPHIC";maintain-aspect-ratio
TRUE;display "USEDEF";valid_file "F";width 3.1324in;height 1.7659in;depth
0pt;original-width 8.8885in;original-height 5.0004in;cropleft "0";croptop
"1";cropright "1";cropbottom "0";filename 'AER
submission/mmexport1663645496832.png';file-properties "XNPEU";}}In this
one-player game (called maze), the player has to enter at the top-left
corner and exit at the bottom-right corner. Suppose that the player can
never turn back, and getting stuck yields a payoff of $0$, whereas reaching
the exit yields a payoff of $1$. At the first decision node, the player can
go left (green arrow) or right (red arrow). Going right strongly obviously
dominates going left. This is because, after going right, the player may
reach the exit or not, depending on her future moves, whereas after going
left the player always gets stuck. However, going right does not locally
dominate going left. This is because, in the unique scenario in which the
game continues after both actions, the continuation game is completely
different after the two actions. Indeed, the choice between the two actions
is not irrelevant for the final outcome.

\subsection{Other related literature}

Our theory of simplicity is related to the literature on limited foresight
and coarse contingent reasoning. While we tackle the issue of foresight in a
very different way than in the literature, our representation of contingent
reasoning is in line with recent contributions on this topic. Like us, Zhang
and Levin (2021)\ model coarse contingent reasoning in games with a
partition of the contingencies. However, Zhang and Levin take a global view
of the game and fix a partition for the entire game exogenously at the
outset, whereas our partitions depend on the information set, and even on
the action pair under comparison. Chew and Wang (2022) only fix a
cardinality $k$ for the partition and stipulate that a strategy $k$%
-dominates another strategy when there exists a partition of cardinality $k$
such that the usual \textquotedblleft worst vs best\textquotedblright\
outcome comparison goes through within each partition element. Their notion
of dominance is motivated by the sure thing principle as first described by
Savage (1954) in his motivating example: if an agent finds that under each
of two complementary events one option is better than the other, then she
should find the former option better before the resolution of the
uncertainty. In contrast, our notion of irrelevance of a scenario is
inspired by Savage's (1954) formal axiom, which says that a decision-maker
should ignore a scenario under which all the alternatives yield the same
outcome. Saponara (2022) considers a decision maker who evaluates each
available act under a specific partition of the contingencies. For each
element of the partition, the act is assigned the minimum attained utility.
The decision maker then computes the expectation of these utilities with
respect to a belief over the partition elements. Following Li (2017),
instead, we adopt a belief-free approach and we require dominance to hold
when the dominated action is assigned the best possible outcome in each
non-irrelevant scenario. Karni and Viero (2013) consider a decision maker
who progressively constructs a state space from the feasible acts she
encounters. In particular, if two acts $f$ and $g$ can give the same two
consequences $c$ and $d$, she will recognize the existence of (at most)\
four possible states: the state where $f$ and $g$ give $c$, the state where
they give $d$, the state where $f$ gives $c$ and $g$ gives $d$, and vice
versa. Let $f$ and $g$ be our two actions under comparison and let $c$ and $%
d $ be the consequence that the game ends or not: the induced state space
coincides with our partition of the true state space.

Going in an opposite direction compared to our work and the literature on
obvious dominance, Borgers and Li (2019) define a more permissive simplicity
standard than strategy-proofness. In particular, they identify a class of
\textquotedblleft simple mechanisms\textquotedblright\ that only depend on
the first-order beliefs about the types of the opponents. Dworczak and Li
(2021)\ relax (obvious) strategy-proofness by allowing for multiplicity of
(obviously) undominated strategies, provided that they all induce the
desired outcome. In this way, they can implement social choice functions
that cannot be implemented in (obviously) dominant strategies. This
relaxation of (obvious)\ strategy-proofness allows, for each profile of
players' types, a multiplicity of undominated behaviors outside of the path
that leads to the desired outcome. This is akin to requiring the existence
of dominant actions for all players only along a path, a notion of
\textquotedblleft on-path strategy-proofness\textquotedblright\ which is
natural in our framework and has some desirable properties: we illustrate it
by example in the Appendix.

In simultaneous and independent works, Bo and Hakimov (2022) and Mackenzie
and Zhou (2022)\ introduce two classes of mechanisms (respectively,
\textquotedblleft pick an object mechanisms\textquotedblright\ and the more
general \textquotedblleft menu mechanisms\textquotedblright ) that encompass
our dynamic TTC\ mechanism as a special case. Bo and\ Hakimov (2022) provide
experimental evidence of the simplicity of a version of their mechanism
(very similar to ours)\ precisely for the TTC\ rule, and justify it
theoretically with a notion of \textquotedblleft robust truthful
equilibrium\textquotedblright ; Mackenzie and Zhou (2022) prove the
existence of versions of dominant-strategy equilibrium in their mechanisms.
It would be interesting to investigate whether some of their mechanisms
yield locally strategy-proof implementation of other social choice functions
than the TTC\ rule. This is far from guaranteed, because a player's moves
may affect her future menus and the opponents' menus; in this case, one
cannot guarantee to a player that her current choice, if it does not yield
an immediate outcome, will not affect the final outcome, making it
impossible to satisfy local dominance.

\section{Appendix}

\textbf{Proof of Proposition \ref{Proposition foundation}. }Weak dominance.
If. Fix $s_{i}\in S_{i}(h,a_{i})$. By assumption, there exists $\overline{s}%
_{i}\in S_{i}(h,\overline{a}_{i})$ that weakly dominates $s_{i}$. Condition (%
\ref{WD1}) implies condition (\ref{LWD1}). Hence, $a_{i}$ is weakly
dominated by $\overline{a}_{i}$.

Only if. Fix $s_{i}\in S_{i}(h,a_{i})$. By assumption, at $h$, $\overline{a}%
_{i}$ weakly dominates $a_{i}$, and thus by Definition \ref{Def:
weak/obvious} there exists $\overline{s}_{i}\in S_{i}(h,\overline{a}_{i})$
such that%
\begin{equation}
\forall s_{-i}\in S_{-i}(h),\text{ \ \ }u_{i}\left( \zeta (\overline{s}%
_{i},s_{-i})\right) \geq u_{i}\left( \zeta (s_{i},s_{-i})\right) .
\label{Eq: theorem 1, 1}
\end{equation}%
Define $\overline{s}_{i}^{\prime }$ as $\overline{s}_{i}^{\prime }(h^{\prime
})=\overline{s}_{i}(h^{\prime })$ if $h^{\prime }\succeq h$ and $\overline{s}%
_{i}^{\prime }(h^{\prime })=s_{i}(h^{\prime })$ if $h^{\prime }\not\succeq h$%
. Thus,%
\begin{eqnarray}
\forall s_{-i} &\in &S_{-i}(h),\text{ \ \ }\zeta (\overline{s}_{i}^{\prime
},s_{-i})=\zeta (\overline{s}_{i},s_{-i}),  \label{Eq: theorem 1, 3} \\
\forall s_{-i} &\not\in &S_{-i}(h),\text{ \ \ }\zeta (\overline{s}%
_{i}^{\prime },s_{-i})=\zeta (s_{i},s_{-i}).  \label{Eq: theorem 1, 4}
\end{eqnarray}%
By (\ref{Eq: theorem 1, 3}), (\ref{Eq: theorem 1, 1}), and (\ref{Eq: theorem
1, 4}), we have%
\begin{equation*}
\forall s_{-i}\in S_{-i},\text{ \ \ }u_{i}\left( \zeta (\overline{s}%
_{i}^{\prime },s_{-i})\right) \geq u_{i}\left( \zeta (s_{i},s_{-i})\right) ,
\end{equation*}%
i.e., condition (\ref{WD1}): $s_{i}$ is weakly dominated by $\overline{s}%
_{i}^{\prime }$.

Obvious dominance. If. Fix $s_{i}\in S_{i}(h,a_{i})$. By assumption, there
exists $\overline{s}_{i}\in S_{i}(h,\overline{a}_{i})$ that obviously
dominates $s_{i}$. Since $s_{i}(h)\not=\overline{s}_{i}(h)$, $h\in \mathcal{D%
}(\overline{s}_{i},s_{i})$. Thus, condition (\ref{OD1}) implies condition (%
\ref{LOD1}). So, $a_{i}$ is obviously dominated by $\overline{a}_{i}$.

Only if. Fix $s_{i}\in S_{i}(h,a_{i})$. By assumption, at $h$, $\overline{a}%
_{i}$ obviously dominates $a_{i}$, and thus by Definition \ref{Def:
weak/obvious} there exists $\overline{s}_{i}\in S_{i}(h,\overline{a}_{i})$
such that%
\begin{equation}
\min_{s_{-i}\in S_{-i}(h)}u_{i}\left( \zeta (\overline{s}_{i},s_{-i})\right)
\geq \max_{s_{-i}\in S_{-i}(h)}u_{i}\left( \zeta (s_{i},s_{-i})\right) .
\label{Eq: prop1, OD1}
\end{equation}%
Define $\overline{s}_{i}^{\prime }$ as $\overline{s}_{i}^{\prime }(h^{\prime
})=\overline{s}_{i}(h^{\prime })$ if $h^{\prime }\succeq h$ and $\overline{s}%
_{i}^{\prime }(h^{\prime })=s_{i}(h^{\prime })$ if $h^{\prime }\not\succeq h$%
. Thus,%
\begin{eqnarray}
\forall s_{-i} &\in &S_{-i}(h),\text{ \ \ }\zeta (\overline{s}_{i}^{\prime
},s_{-i})=\zeta (\overline{s}_{i},s_{-i}),  \label{Eq: Prop1, OD2} \\
\forall s_{-i} &\not\in &S_{-i}(h),\text{ \ \ }\zeta (\overline{s}%
_{i}^{\prime },s_{-i})=\zeta (s_{i},s_{-i}).  \label{Eq: Prop1, OD3}
\end{eqnarray}%
By (\ref{Eq: Prop1, OD2}) and (\ref{Eq: prop1, OD1}), we have%
\begin{equation*}
\min_{s_{-i}\in S_{-i}(h)}u_{i}\left( \zeta (\overline{s}_{i}^{\prime
},s_{-i})\right) \geq \max_{s_{-i}\in S_{-i}(h)}u_{i}\left( \zeta
(s_{i},s_{-i})\right) ,
\end{equation*}%
that is, $\overline{s}_{i}^{\prime }$ and $s_{i}$ satisfy condition (\ref%
{OD1}) at $h$.\ Since $s_{i},\overline{s}_{i}^{\prime }\in S_{i}(h)$ but $%
\overline{s}_{i}^{\prime }(h)\not=s_{i}(h)$, we have $h\in \mathcal{D}(%
\overline{s}_{i},s_{i})$. By (\ref{Eq: Prop1, OD3}), there is no point of
departure between $s_{i}$ and $\overline{s}_{i}^{\prime }$ along any path
that does not go through $h$. So, $\mathcal{D}(\overline{s}%
_{i},s_{i})=\left\{ h\right\} $. Thus, $s_{i}$ is obviously dominated by $%
\overline{s}_{i}^{\prime }$. $\blacksquare $

\bigskip

\textbf{Proof of Theorem \ref{CHAR 1}. }Weak dominance. Only if. Fix $%
\overline{h}\in H_{i}^{\ast }(\overline{s}_{i})$. Fix $a_{i}\in A_{i}^{%
\overline{h}}\backslash \left\{ \overline{s}_{i}(\overline{h})\right\} $.
For every $s_{i}\in S_{i}(\overline{h},a_{i})$, since $\overline{s}_{i}\in
S_{i}(\overline{h},\overline{s}_{i}(\overline{h}))$ is weakly dominant, it
weakly dominates $s_{i}$. Hence, by Proposition \ref{Proposition foundation}
(if part) $\overline{s}_{i}(\overline{h})$ weakly dominates $a_{i}$.

If. Let $\overline{s}_{i}$ be a strategy that prescribes a weakly dominant
action at every $h\in H_{i}^{\ast }(\overline{s}_{i})$. Fix $s_{-i}\in
S_{-i} $. Fix $\overline{h}\prec \zeta (\overline{s}_{i},s_{-i})$ and
suppose by way of induction that, for every $h^{\prime }\in H_{i}^{\ast }$
such that $\overline{h}\prec h^{\prime }\prec \zeta (\overline{s}%
_{i},s_{-i}) $,\footnote{%
If $\overline{h}$ is a stage-$T$ information set, or anyway the last active
information set along path $\zeta (\overline{s}_{i},s_{-i})$, the induction
hypothesis is vacuously satisfied.} for every $s_{i}^{\prime }$ that departs
from $\overline{s}_{i}$ at $h^{\prime }$,%
\begin{equation*}
u_{i}\left( \zeta (\overline{s}_{i},s_{-i})\right) \geq u_{i}\left( \zeta
(s_{i}^{\prime },s_{-i})\right) .
\end{equation*}%
Fix $s_{i}$ that departs from $\overline{s}_{i}$ at $\overline{h}$. By
Proposition \ref{Proposition foundation} (only if part) there exists $%
\overline{s}_{i}^{\prime }\in S_{i}(\overline{h},\overline{s}_{i}(\overline{h%
}))$ that weakly dominates $s_{i}$, and thus%
\begin{equation}
u_{i}\left( \zeta (\overline{s}_{i}^{\prime },s_{-i})\right) \geq
u_{i}\left( \zeta (s_{i},s_{-i})\right) .  \label{Eq: th1, 1}
\end{equation}%
Either $\zeta (\overline{s}_{i},s_{-i})=\zeta (\overline{s}_{i}^{\prime
},s_{-i})$, or $\overline{s}_{i}^{\prime }$ departs from $\overline{s}_{i}$
at some $h^{\prime }\in H_{i}^{\ast }$ such that $\overline{h}\prec
h^{\prime }\prec \zeta (\overline{s}_{i},s_{-i})$, and also in this second
case, by the induction hypothesis,%
\begin{equation}
u_{i}\left( \zeta (\overline{s}_{i},s_{-i})\right) \geq u_{i}\left( \zeta (%
\overline{s}_{i}^{\prime },s_{-i})\right) .  \label{Eq: th1, 2}
\end{equation}%
Inequalities (\ref{Eq: th1, 2}) and (\ref{Eq: th1, 1}) yield $u_{i}\left(
\zeta (\overline{s}_{i},s_{-i})\right) \geq u_{i}\left( \zeta
(s_{i},s_{-i})\right) $. Clearly, the same holds (with equality) also for
all $s_{i}\in S_{i}$ that do not depart from $\overline{s}_{i}$ at any $%
h\prec \zeta (\overline{s}_{i},s_{-i})$. Thus, since $s_{-i}$ was arbitrary, 
$\overline{s}_{i}$ weakly dominates every $s_{i}\in S_{i}$.

Obvious dominance. Only if. Fix $\overline{h}\in H_{i}^{\ast }(\overline{s}%
_{i})$. Fix $a_{i}\in A_{i}^{\overline{h}}\backslash \left\{ \overline{s}%
_{i}(\overline{h})\right\} $. For every $s_{i}\in S_{i}(\overline{h},a_{i})$%
, since $\overline{s}_{i}\in S_{i}(\overline{h},\overline{s}_{i}(\overline{h}%
))$ is obviously dominant, it obviously dominates $s_{i}$. Hence by
Proposition \ref{Proposition foundation} (if part) $\overline{s}_{i}(%
\overline{h})$ obviously dominates $a_{i}$.

If. Let $\overline{s}_{i}$ be a strategy that prescribes an obviously
dominant action at every $h\in H_{i}^{\ast }(\overline{s}_{i})$. Fix $%
\overline{h}\in H_{i}^{\ast }(\overline{s}_{i})$ and suppose by way of
induction that for every $h^{\prime }\in H_{i}^{\ast }(\overline{s}_{i})$
that follows $\overline{h}$,\footnote{%
When no such $h^{\prime }$ exists, the induction hypothesis is vacuously
satisfied.} for every $s_{i}^{\prime }$ that departs from $\overline{s}_{i}$
at $h^{\prime }$,%
\begin{equation*}
\min_{s_{-i}\in S_{-i}(h^{\prime })}u_{i}\left( \zeta (\overline{s}%
_{i},s_{-i})\right) \geq \max_{s_{-i}\in S_{-i}(h^{\prime })}u_{i}\left(
\zeta (s_{i}^{\prime },s_{-i})\right) .
\end{equation*}%
Fix $s_{i}$ that departs from $\overline{s}_{i}$ at $\overline{h}$. By
Proposition \ref{Proposition foundation} (only if part) there exists $%
\overline{s}_{i}^{\prime }\in S_{i}(\overline{h},\overline{s}_{i}(\overline{h%
}))$ that obviously dominates $s_{i}$, and thus%
\begin{equation}
\min_{s_{-i}\in S_{-i}(\overline{h})}u_{i}\left( \zeta (\overline{s}%
_{i}^{\prime },s_{-i})\right) \geq \max_{s_{-i}\in S_{-i}(\overline{h}%
)}u_{i}\left( \zeta (s_{i},s_{-i})\right) .  \label{Th 1, OD1}
\end{equation}%
For each $s_{-i}\in S_{-i}(\overline{h})$, either $\zeta (\overline{s}%
_{i},s_{-i})=\zeta (\overline{s}_{i}^{\prime },s_{-i})$, or $\overline{s}%
_{i}^{\prime }$ departs from $\overline{s}_{i}$ at some $h^{\prime }\in
H_{i}^{\ast }$ such that $\overline{h}\prec h^{\prime }\prec \zeta (%
\overline{s}_{i},s_{-i})$, and also in this second case, by the induction
hypothesis,%
\begin{equation}
u_{i}\left( \zeta (\overline{s}_{i},s_{-i})\right) \geq u_{i}\left( \zeta (%
\overline{s}_{i}^{\prime },s_{-i})\right) \text{.}  \label{Th 1, OD2}
\end{equation}%
Inequality (\ref{Th 1, OD2}) for all $s_{-i}\in S_{-i}(\overline{h})$, along
with inequality (\ref{Th 1, OD1}), yield%
\begin{equation*}
\min_{s_{-i}\in S_{-i}(\overline{h})}u_{i}\left( \zeta (\overline{s}%
_{i},s_{-i})\right) \geq \max_{s_{-i}\in S_{-i}(\overline{h})}u_{i}\left(
\zeta (s_{i},s_{-i})\right) .
\end{equation*}%
Since this holds for all $\overline{h}\in \mathcal{D}(\overline{s}%
_{i},s_{i}) $, $\overline{s}_{i}$ obviously dominates $s_{i}$. $\blacksquare 
$

\bigskip

\textbf{Proof of Theorem \ref{Theorem local dominant = local c-dominant}. }%
Only if: by inspection of the definitions.

If. Let $\overline{s}_{i}$ be a strategy that prescribes an s-dominant
action at every $h\in H_{i}^{\ast }(\overline{s}_{i})$. First we prove that $%
\overline{s}_{i}$ is weakly dominant. Fix $\overline{h}\in H_{i}^{\ast }(%
\overline{s}_{i})$ and suppose by way of induction that for every $h\in
H_{i}^{\ast }(\overline{s}_{i})$ that follows $\overline{h}$,%
\footnotemark[35]%
\begin{equation}
\forall s_{i}\in S_{i}(h),\forall s_{-i}\in S_{-i}(h),\text{ \ \ }%
u_{i}\left( \zeta (\overline{s}_{i},s_{-i})\right) \geq u_{i}\left( \zeta
(s_{i},s_{-i})\right) .  \label{Th 2 IH}
\end{equation}%
Fix $s_{i}\in S_{i}(\overline{h})$ and $s_{-i}\in S_{-i}(\overline{h})$.
Since $\overline{s}_{i}(\overline{h})$ is s-dominant, it is wishfully
dominant, and thus there exists $\overline{s}_{i}^{\prime }\in S_{i}(%
\overline{h},\overline{s}_{i}(\overline{h}))$ such that%
\begin{equation*}
u_{i}\left( \zeta (\overline{s}_{i}^{\prime },s_{-i})\right) \geq
u_{i}\left( \zeta (s_{i},s_{-i})\right) .
\end{equation*}%
Since $\overline{s}_{i}^{\prime }(\overline{h})=\overline{s}_{i}(\overline{h}%
)$, either $\zeta (\overline{s}_{i},s_{-i})=\zeta (\overline{s}_{i}^{\prime
},s_{-i})$, or there exists $h^{\prime }\in H_{i}^{\ast }(\overline{s}_{i})$
that follows $\overline{h}$ such that $\overline{s}_{i}^{\prime }\in
S_{i}(h^{\prime })$ and $s_{-i}\in S_{-i}(h^{\prime })$, and also in this
second case, by the induction hypothesis,%
\begin{equation*}
u_{i}\left( \zeta (\overline{s}_{i},s_{-i})\right) \geq u_{i}\left( \zeta (%
\overline{s}_{i}^{\prime },s_{-i})\right) .
\end{equation*}%
So we have $u_{i}\left( \zeta (\overline{s}_{i},s_{-i})\right) \geq
u_{i}\left( \zeta (s_{i},s_{-i})\right) $, as desired. Thus, for every $%
s_{i}\in S_{i}$, for each $s_{-i}\in S_{-i}$, if there is $h\in H_{i}^{\ast
}(\overline{s}_{i})$ such that $s_{i}\in S_{i}(h)$ and $s_{-i}\in S_{-i}(h)$%
, then we have just shown that $u_{i}\left( \zeta (\overline{s}%
_{i},s_{-i})\right) \geq u_{i}\left( \zeta (s_{i},s_{-i})\right) $,
otherwise, $\zeta (\overline{s}_{i},s_{-i})=\zeta (s_{i},s_{-i})$. Hence, $%
\overline{s}_{i}$ weakly dominates $s_{i}$.

Now, we prove that, for each $h\in H_{i}^{\ast }(\overline{s}_{i})$ and $%
a_{i}\in A_{i}(h)\backslash \left\{ \overline{s}_{i}(h)\right\} $, $%
\overline{s}_{i}(h)$ dominates $a_{i}$ given the partition $\mathcal{S}^{h}$
of $S_{-i}(h)$ that player $i$ uses for the comparison (both with
s-dominance and dominance). For every $s_{i}\in S_{i}(h,a_{i})$ and $%
\overline{S}_{-i}\in \mathcal{S}^{h}$, by condition (\ref{LCWD1}), there
exists $\overline{s}_{i}^{\prime }\in S_{i}(h)$ such that%
\begin{equation*}
\min_{s_{-i}\in \overline{S}_{-i}}u_{i}\left( \zeta (\overline{s}%
_{i}^{\prime },s_{-i})\right) \geq \max_{s_{-i}\in \overline{S}%
_{-i}}u_{i}\left( \zeta (s_{i},s_{-i})\right) \text{.}
\end{equation*}%
Since $\overline{s}_{i}$ is weakly dominant,%
\begin{equation*}
\min_{s_{-i}\in \overline{S}_{-i}}u_{i}\left( \zeta (\overline{s}%
_{i},s_{-i})\right) \geq \min_{s_{-i}\in \overline{S}_{-i}}u_{i}\left( \zeta
(\overline{s}_{i}^{\prime },s_{-i})\right) .
\end{equation*}%
The last two inequalities combined yield condition (\ref{LD1}): $\overline{s}%
_{i}(h)$ dominates $a_{i}$. $\blacksquare $

\bigskip

\textbf{Proof of Proposition \ref{Prop: auction}. }Fix an information set $h$
where bidder $i$ has not left the auction yet. First, we compare $b$ and $%
\ell $, regardless of the presence of $w$. The partition $\mathcal{S}^{\ell
}(h,b,\ell )$ features only two non-empty scenarios:\ the scenario $%
S^{b,\ell }(h)$ where none of the opponents bid at round $t$, and the
scenario $S_{b}^{\ell }(h)$ where at least one does --- $S_{b,\ell }(h)$ and 
$S_{\ell }^{b}(h)$ are empty because after $\ell $ there is no active
information set of player $i$. We now check local dominance. In scenario $%
S^{b,\ell }(h)$, the only possible outcome after $b$ is that $i$ wins the
auction, therefore either $b$, if $t\leq v-1$, or $\ell $, if $t\geq v$,
satisfies condition (\ref{VLDbis}) (with the role of $\overline{a}_{i}$).
For the scenario $S_{b}^{\ell }(h)$, consider first $t\geq v$. In this case, 
$\ell $ terminates the game with payoff $0$, while after $b$ the payoff
cannot be strictly positive, therefore $\ell $ satisfies condition (\ref%
{VLDbisbis}) (with $\overline{a}_{i}=\ell $). Thus, $\ell $ locally
dominates $b$. Consider now $t\leq v-1$. The strategy that prescribes $\ell $
at every information set after $b$ mimicks the dummy continuation strategy
that prescribes $\ell $ at every information set after $\ell $. With the
mimicking continuation strategy (in the scenario under consideration), $i$'s
payoff will be determined at stage $t+1$, and since $t+1\leq v$, it cannot
be negative. Player $i$' payoff after $\ell $ is always $0$ in scenario $%
S_{\ell }^{b}(h)$. Therefore, condition (\ref{VLDtris}) holds. Thus, $b$
locally dominates $\ell $.

Now we compare $b$ with $w$ for $t\leq v-1$. If only one opponent is still
in the auction, $w$ is equivalent to $\ell $, because it always terminates
the auction. Otherwise, the partition $\mathcal{S}^{\ell }(h,b,w)$ features
three non-empty scenarios: the scenario $S^{b,w}(h)$ where none of the
opponents bid at round $t$, the scenario $S_{b}^{w}(h)$ where only one bids,
and the scenario $S_{b,w}(h)$ where more than one bids --- $S_{w}^{b}(h)$ is
empty because if the auction ends with $b$, so it does with $w$. For the
first two scenarios, the comparison between $b$ and $w$ is identical to the
comparison between $b$ and $\ell $. Note in particular that in scenario $%
S_{b}^{w}(h)$, player $i$ is forced to play $\ell $ after $w$, therefore the
mimicking continuation strategy after $b$ prescribes $\ell $ as before. For
scenario $S_{b,w}(h)$, we show that it is irrelevant, so that, as for the
comparison with $\ell $, $b$ locally dominates $w$. Thus, for $t<v$ let $(%
\overline{a}_{i},a_{i})=(b,w)$, for $t\geq v$ let $(\overline{a}%
_{i},a_{i})=(w,b)$. Fix $s_{i}\in S_{i}(h,a_{i})$. Construct the mimicking $%
\overline{s}_{i}\in S_{i}(h,\overline{a}_{i})$ as follows. Fix a stage $%
t^{\prime }>t$ and suppose by way of induction that, for each $s_{-i}\in
S_{b,w}(h)$, either the object was assigned before stage $t^{\prime }$ under
both $s_{i}$ and the $\overline{s}_{i}$ under construction, or it was not
assigned before stage $t^{\prime }$ under both $s_{i}$ and the $\overline{s}%
_{i}$ under construction. Under all $s_{-i}\in S_{b,w}(h)$ that fall into
the first category, $s_{i}$ prescribes $\ell $ at the stage-$t^{\prime }$
information set $h^{\prime }\prec \zeta (s_{i},s_{-i})$, and we can let $%
\overline{s}_{i}$ prescribe $\ell $ as well, at every stage-$t^{\prime }$
information set that is consistent with $\overline{s}_{i}$ and with the fact
that the auction ended. Under all $s_{-i}\in S_{b,w}(h)$ that fall into the
second category, player $i$ reaches the same stage-$t^{\prime }$ information
set $h^{\prime }$ with $s_{i}$, and the same stage-$t^{\prime }$ information
set $\overline{h}^{\prime }$ with $\overline{s}_{i}$. This is because all
that player $i$ learnt in the previous stages is that the auction did not
end.\footnote{%
To be precise, regarding the behavior of the opponents at stage $t$, after $%
b $ player $i$ has less accurate information than after $w$, because for the
auction to continue after waiting it takes at least two opponents to bid at
round $t$ instead of one. Nonetheless, \emph{in the scenario where the
auction continues after both actions}, there is just one information set per
stage also after $w$.} Since the auction is still ongoing both at $h^{\prime
}$ and $\overline{h}^{\prime }$, all actions are available and we can let $%
\overline{s}_{i}(\overline{h}^{\prime })=s_{i}(h^{\prime })$. Moreover, at
stage $t^{\prime }$, the opponents reach the same information set regardless
of $s_{i}$ or $\overline{s}_{i}$, because all they observe is that the
auction is still ongoing.\footnote{%
Given that under $s_{-i}$ the choice of $i$ at $h$ is not pivotal to
determine whether the auction continues, it means that $i$'s opponents
cannot infer her choice at $h$ from the fact that the auction continued.}
Hence, just like player $i$, they make the same move at stage $t^{\prime }$
regardless of $s_{i}$ or $\overline{s}_{i}$. So, at stage $t^{\prime }$, if
the auction ends under $s_{i}$ it also ends under $\overline{s}_{i}$, and if
it continues under $s_{i}$, it also continues under $\overline{s}_{i}$: the
induction hypothesis for stage $t^{\prime }+1$ is proven.\ Moreover, by the
same token, if the auction ends at stage $t^{\prime }$, the outcome is the
same under $s_{i}$ and $\overline{s}_{i}$. Thus, irrelevance holds.

To conclude the proof, we need to show that $\ell $ locally dominates $w$
when $t=v$. There two non-empty scenarios: $S^{\ell ,w}(h)$ and $S_{w}^{\ell
}(h)$. In the first scenario, since $t=v$, both actions always yield payoff $%
0$, thus $\ell $ satisfies condition (\ref{VLDbis}) (with the role of $%
\overline{a}_{i}$). In the second scenario, $\ell $ terminates the game with
payoff $0$, while after $w$ the payoff cannot be strictly positive,
therefore $\ell $ satisfies condition (\ref{VLDbisbis}) (with $\overline{a}%
_{i}=\ell $). $\blacksquare $

\bigskip

\textbf{Proof of Theorem \ref{Theorem: TTC}. }Fix a player $i\in I$, a stage 
$t\in \left\{ 1,...,N\right\} $, and a stage-$t$ information set $h\in
H_{i}^{\ast }$. Let $a$ be $i$'s favorite still-available object, and let $b$
be another available objects. Thus, $a,b\in A_{i}^{h}$. We are going to show
that $a$ locally dominates $b$.

Note that the object assigned to a player coincides with the last object she
names. Therefore, in the scenarios $S^{a,b}(h)$ and $S_{b}^{a}(h)$, player $%
i $ gets $a$ after naming $a$, so Conditions (\ref{VLDbis}) and (\ref%
{VLDbisbis}) are satisfied.

We will show that, for every $s_{i}\in S_{i}(h,b)$, there exists $\overline{s%
}_{i}\in S_{i}(h,a)$ that mimics $s_{i}$ under $S_{a}^{b}(h)\cup S_{a,b}(h)$%
. For every $s_{-i}\in S_{a}^{b}(h)\cup S_{a,b}(h)$, player $i$ moves again
after choosing $a$ at $h$, therefore $\overline{s}_{i}$ and $s_{i}$
prescribe the same last move along the paths induced by $(s_{i},s_{-i})$ and
by $(\overline{s}_{i},s_{-i})$, and hence yield the same outcome. Thus, $%
S_{a,b}(h)$ is irrelevant. Moreover, given that $s_{i}$ yields $b$ under all 
$s_{-i}\in S_{a}^{b}$, so does $\overline{s}_{i}$, and condition (\ref%
{VLDtris}) is satisfied as an equality.

Now we construct the mimicking $\overline{s}_{i}$. We will repeatedly use
the fact that players only observe the history of available objects,
therefore as long as player $i$ does not obtain an object, her moves cannot
affect the moves of the opponents, and hence the history of available items.
Fix $h^{\prime }\in \mathcal{H}^{\ast }(h,a)$. Let $\psi $ denote the
history of available objects that player $i$ observes at $h^{\prime }$.
First we prove the following fact

\begin{claim}
\label{Claim: available}For each $s_{-i}\in S_{-i}(h^{\prime })$, if under $%
s_{i}$ player $i$ is assigned an object by stage $\tau (h^{\prime })$, then
that object is available at $h^{\prime }$.
\end{claim}

\textbf{Proof of the claim. }Suppose that, under $s_{-i}$ and $s_{i}$,
player $i$ is assigned an object $c$ at some stage $t^{\prime }\leq \tau
(h^{\prime })$. Then, until stage $t^{\prime }$, under $s_{-i}$, the moves
of the opponents are the same under $s_{i}$ and after choosing $a$ at $h$.
Therefore, also after choosing $a$ at $h$, at stage $t^{\prime }$ the owner
of $c$ is pointing, directly or indirectly, to player $i$. Since this player
and possibly the other players in the chain that links them to player $i$
cannot move as long as player $i$ is not assigned an object, $c$ remains
available at\ $h^{\prime }$. $\square $

Now determine $\overline{s}_{i}(h^{\prime })$ as follows. For all $s_{-i}\in
S_{-i}(h^{\prime })$, as long as player $i$ does not get an object, the
history of available objects follows $\psi $ also under $s_{i}$. Thus, if $b$
is available at $h^{\prime }$, under $s_{i}$ and all $s_{-i}\in
S_{-i}(h^{\prime })$ either player $i$ gets $b$ by stage $\tau (h^{\prime })$%
, or she cannot move until stage $\tau (h^{\prime })$; either way, she will
be naming $b$ at stage $\tau (h^{\prime })$, so we can let $\overline{s}%
_{i}(h^{\prime })=b$. If instead object $b$ stops being available along $%
\psi $ at some stage $t^{\prime }\leq \tau (h^{\prime })$, by Claim \ref%
{Claim: available} player $i$ did not obtain $b$ by stage $t^{\prime }$
under any $s_{-i}\in S_{-i}(h^{\prime })$. But then, using $s_{i}$ and $\psi 
$, we can determine the \emph{unique} object $c$ player $i$ names at stage $%
t^{\prime }$ under $s_{i}$ and \emph{all} $s_{-i}\in S_{-i}(h^{\prime })$.
Repeating the reasoning we made for $b$, we can determine if player $i$
names $c$ at stage $\tau (h^{\prime })$, or will switch to another object $d$
at some stage $t^{\prime \prime }\leq \tau (h^{\prime })$. Going on in this
fashion, we can determine the \emph{unique} object $e$ player $i$ names at
stage $\tau (h^{\prime })$ under $s_{i}$ and \emph{all} $s_{-i}\in
S_{-i}(h^{\prime })$, and let $\overline{s}_{i}(h^{\prime })=e$.

Now fix a stage $t^{\prime \prime }>\tau (h^{\prime })$, a stage-$t^{\prime
\prime }$ information set $h^{\prime \prime }\succ h^{\prime }$, and suppose
by way of induction that $\overline{s}_{i}$ was defined and mimics $s_{i}$
at every $\widetilde{h}\prec h^{\prime \prime }$. Thus, for all $s_{-i}\in
S_{-i}(h^{\prime \prime })$, under both $s_{i}$ and $\overline{s}_{i}$,
either before stage $\tau (h^{\prime \prime })$ player $i$ has already
obtained the same object $e$, and then we can let $\overline{s}%
_{i}(h^{\prime \prime })=e$, or player $i$ has not obtained any object. In
this second case, the history of available objects at $h^{\prime \prime }$
coincides with the one at the\emph{\ unique} stage-$\tau (h^{\prime \prime
}) $ information set $\widetilde{h}$ that is consistent with $s_{i}$ and 
\emph{all} $s_{-i}\in S_{-i}(h^{\prime \prime })$, and this also implies
that $h^{\prime \prime }\in H_{i}^{\ast }$ if and only if $\widetilde{h}\in
H_{i}^{\ast }$. So, we can let $\overline{s}_{i}(h^{\prime \prime })=s_{i}(%
\widetilde{h})$.

To conclude, note that $\cup _{h^{\prime }\in \mathcal{H}^{\ast
}(h,a)}S_{-i}(h^{\prime })=S_{a}^{b}(h)\cup S_{a,b}(h)$, so $\overline{s}%
_{i} $ mimics $s_{i}$ under $S_{a}^{b}(h)\cup S_{a,b}(h)$. $\blacksquare $

\bigskip

\subsection*{Example of on-path strategy proofness}

A shepherd dog has to recall the sheep from the top of the hill for the
night. The dog's goal is to maximize the number of sheep that make it all
the way down to the sheepfold before falling asleep. Then, by contract, the
dog has to guard the sheep from $2/3$ of their average sleeping altitude.%
\footnote{%
This game is a dynamic transformation of \textquotedblleft guess $2/3$ of
the average\textquotedblright , which obviates the possible distrust in the
opponents' rationality by letting players \emph{observe} whether the
opponents have played rationally. Glazer and Rubinstein (1996) provide
general rules to transform a dominance-solvable static game into a
strategy-proof dynamic one.} A sheep's payoff is the distance from the dog
during the sleep. Because of the fog, the sheep cannot see where they are
going and typically sleep scattered on the slope. To solve this problem, the
dog comes up with the following idea. The dog first stands at altitude $67$
(the top of the hill is at altitude $100$) and shines a light. The sheep can
see the light through the fog and walk down towards it. Those who stop along
the way (at an altitude between $68$ and $100$) cannot help falling asleep.
Those who reach the dog at $67$ get to see each other and manage to stay
awake. Then the dog moves to altitude $45$ and shines a light again for the
sheep that are still awake. The game continues in this fashion until the dog
reaches the sheepfold at altitude $1$. Then, if some sheep has not arrived,
the dog looks for his prescribed guarding position and stops there.

\bigskip

Take the viewpoint of a sheep at the initial history or at an information
set where it has reached the dog and sees that all other sheep have reached
the dog as well. Reaching the dog at the next, lower altitude is locally
dominant and obviously dominant. Given any alternative action of stopping at
a higher altitude, the local partition coincides with the trivial partition,
as there is only one scenario: the one in which the alternative action
terminates the game (the sheep then falls asleep) and reaching the dog does
not (unless the dog has reached the sheepfold, in which case the game ends
also after reaching the dog). The sheep can entertain the continuation
strategy (after reaching the dog) of not moving forward and sleep there.
Such a continuation strategy mimicks the dummy continuation strategy after
the alternative action. If the sheep reaches the dog and sleeps there, the
payoff will certainly be higher than after the alternative action:\ given
that all sheep have reached the dog at its old altitude, the dog will not
sleep below its new altitude, which is $2/3$ of the old one.

\bigskip

Note that the sheep do not have an obviously dominant strategy, or a
strategy entirely made of locally dominant actions. If at some point a sheep
observes that not all others have reached the dog, walking down to the next
dog's altitude might not be optimal, because the dog might have to finally
guard the sheep from a higher altitude. Therefore, the game is not obviously
strategy proof or locally strategy-proof. However, it is \emph{on-path
strategy proof} with both obvious and local dominance, in the sense that
reaching the dog is dominant when all sheep have reached the dog, i.e., when
they have \emph{all} played their dominant action at the previous
information sets. The off-path information sets, though, could be eliminated
from the game: after observing that not all sheep have arrived, the dog
could quit the game and move directly to its guarding position, instead of
trying to drag the remaining sheep down to the sheepfold. Quitting the game
in this way, however, requires the ability of the dog to commit to a
suboptimal behavior given its objective function.


\bigskip

\begin{thebibliography}{99}
\bibitem{BH} \textsc{Bo, I. and R. Hakimov (2020): }\textquotedblleft
Pick-an-object Mechanisms\textquotedblright , working paper.

\bibitem{NS} \textsc{Borgers, T. and J. Li (2019):} \textquotedblleft
Strategically simple mechanisms\textquotedblright , \textit{Econometrica},
87, 2003-2035.

\bibitem{CA} \textsc{Chew, S. H. and W. Wang (2022): }\textquotedblleft
Generalizing obvious dominance using the sure-thing
principle\textquotedblright , working paper.

\bibitem{CHO} \textsc{Dworczak, P., and J. Li (2022):} \textquotedblleft Are
Simple Mechanisms Optimal when Agents are Unsophisticated?\textquotedblright
, working paper.

\bibitem{KHL} \textsc{Kagel, J., Harstad, R., Levin, D. (1987): }%
\textquotedblleft Information impact and allocation rules in auctions with
affiliated private values: a laboratory study\textquotedblright , \textit{%
Econometrica}, 55, 1275-1304.

\bibitem{KV} \textsc{Karni, E. and L-M Viero (2013):} \textquotedblleft
Reverse Bayesianism: A Choice-Based Theory of Growing
Awareness\textquotedblright , \textit{American Economic Review}, 103,
2790-2810.

\bibitem{K} \textsc{Li, S. (2017): }\textquotedblleft Obviously
Strategy-Proof Mechanisms\textquotedblright ,\ \textit{American Economic
Review}, 107, 307-352, 3257-87.

\bibitem{LIU} \textsc{Mackenzie, A. and Y. Zhou (2022): }\textquotedblleft
Menu mechanisms\textquotedblright , \textit{Journal of Economic Theory},
204, 105511.

\bibitem{MA} \textsc{Marx, L. and J. Swinkels (1997): }\textquotedblleft
Order Independence for Iterated Weak Dominance\textquotedblright , \textit{%
Games and Economic Behavior}, 18, 219-245.

\bibitem{P} \textsc{Pycia, M. and P. Troyan (2022): }\textquotedblleft A
Theory of Simplicity in Games and Mechanism Design\textquotedblright ,
working paper.

\bibitem{SAP} \textsc{Saponara, N. (2022):.} \textquotedblleft Revealed
reasoning\textquotedblright , \textit{Journal of Economic Theory}, 199,
105096.

\bibitem{SAV} \textsc{Savage, L. J. (1954):} \textquotedblleft The
foundations of statistics\textquotedblright , John Wiley \& Sons Inc., New
York.

\bibitem{SA} \textsc{Shimoji, M. and J. Watson (1998): }\textquotedblleft
Conditional Dominance, Rationalizability, and Game Forms\textquotedblright , 
\textit{Journal of Economic Theory}, 83, 161-195.

\bibitem{T} \textsc{Zhang, L. and D. Levin (2021): }\textquotedblleft
Partition Obvious Preference and Mechanism Design: Theory and
Experiment\textquotedblright , working paper.
\end{thebibliography}
\end{document}